\documentclass[a4paper,11pt]{article}

\usepackage{jcappub} 
                    
\usepackage[T1]{fontenc} 
\usepackage{lmodern} 
\usepackage{tikz}
\usepackage{float}
\usepackage{fontawesome}
\usepackage{orcidlink}

\usepackage{scalerel,tikz}
\usetikzlibrary{svg.path}
\definecolor{orcidlogocol}{HTML}{A6CE39}
\tikzset{orcidlogo/.pic={
 \fill[orcidlogocol] svg{M256,128c0,70.7-57.3,128-128,128C57.3,256,0,198.7,0,128C0,57.3,57.3,0,128,0C198.7,0,256,57.3,256,128z};
 \fill[white] svg{M86.3,186.2H70.9V79.1h15.4v48.4V186.2z}
 svg{M108.9,79.1h41.6c39.6,0,57,28.3,57,53.6c0,27.5-21.5,53.6-56.8,53.6h-41.8V79.1z M124.3,172.4h24.5c34.9,0,42.9-26.5,42.9-39.7c0-21.5-13.7-39.7-43.7-39.7h-23.7V172.4z}
 svg{M88.7,56.8c0,5.5-4.5,10.1-10.1,10.1c-5.6,0-10.1-4.6-10.1-10.1c0-5.6,4.5-10.1,10.1-10.1C84.2,46.7,88.7,51.3,88.7,56.8z};
}}
\newcommand\orcidicon[1]{\href{https://orcid.org/#1}{\mbox{\scalerel*{
\begin{tikzpicture}[yscale=-1,transform shape]
\pic{orcidlogo};
\end{tikzpicture}
}{|}}}}

\usetikzlibrary{arrows.meta}

\DeclareUnicodeCharacter{2009}{\,}

\title{Simulation budgeting for hybrid effective field theories}

\author[a,b,c]{{Alexa Bartlett,}\orcidlink{0000-0002-3421-8724}}
\author[d,a,b,c]{{Joseph DeRose,}\orcidlink{0000-0002-0728-0960}}
\author[a,b,c]{{and Martin White}\orcidlink{0000-0001-9912-5070}}

\affiliation[a]{Department of Physics, University of California, Berkeley, CA 94720, USA}
\affiliation[b]{Berkeley Center for Cosmological Physics, UC Berkeley, CA 94720, USA}
\affiliation[c]{Lawrence Berkeley National Laboratory, One Cyclotron Road, Berkeley, CA 94720, USA}
\affiliation[d]{Physics Department, Brookhaven National Laboratory, Upton, NY 11973, USA}

\emailAdd{alexa\_bartlett@berkeley.edu}
\emailAdd{jderose@bnl.gov}
\emailAdd{mwhite@berkeley.edu}

\abstract{In this work, we forecast the number of, and requirements on, N-body simulations needed to train hybrid effective field theory (HEFT) emulators for a range of use cases, using a hybrid of {\tt HMcode} and perturbation theory as a surrogate model. Our accuracy goals, determined with careful consideration of statistical and systematic uncertainties, are $1\%$ accurate in the high-likelihood range of cosmological parameters, and $2\%$ accurate over a broader parameter space volume for  $k<1\,h\,{\rm Mpc}^{-1}$ and $z<3$. Focusing in part on the 8-parameter $w_0w_a$CDM+$m_\nu$ cosmological model, we find that $<225$ simulations are required to meet our error goals over our wide parameter space, including models with rapidly evolving dark energy, given our simulation and emulator recommendations. For a more restricted parameter space volume, as few as 80 simulations are sufficient. We additionally present simulation forecasts for example use cases, and make the code used in our analyses publicly available. These results offer practical guidance for efficient emulator design and simulation budgeting in future cosmological analyses.}

\begin{document}
\maketitle
\flushbottom

\section{Introduction}
\label{sec:intro}
Ongoing and future ground- and space-based surveys, including the Nancy Grace Roman Space Telescope (RST) \cite{spergel2015widefieldinfrarredsurveytelescopeastrophysics}, the SPHEREx mission \cite{doré2015cosmologyspherexallskyspectral}, the Euclid satellite \cite{laureijs2011eucliddefinitionstudyreport, Euclid2018}, the Vera Rubin Observatory's Legacy Survey of Space and Time (LSST) \cite{Ivezi__2019}, and the Dark Energy Spectroscopic Instrument (DESI) \cite{desicollaboration2016desiexperimentisciencetargeting, desicollaboration2016desiexperimentiiinstrument, DESI_Collaboration_2022} will provide excellent measurements of large-scale structure (LSS) probes. Using this data to the fullest extent will require accurately modeling multiple cosmological probes and their associated systematics on a wide range of scales. Combinations of analytical calculations, numerical simulations, and machine learning tools will allow us to optimize such modeling and assess when our modeling precision is sufficient for future data. Small-scale clustering and lensing data contain a wealth of cosmological information, but analysis of this data requires accurate models of nonlinear growth of structure, nonlinear galaxy bias, and the effects of baryons. In this paper, we aim to forecast the costs of simulations for simulation-based models and to design a strategy for optimally handling said simulations.

Hybrid effective field theory (HEFT) \cite{Kokron_2021, Modi_2020, Hadzhiyska_2021, Zennaro_2022, Pellejero_Iba_ez_2023, Nicola2023} combines perturbation theory and N-body simulations, allowing us to reap the benefits of both. HEFT is being widely adopted and generalized, and we focus in part on extending and validating this formalism for cosmologies beyond $\Lambda$CDM, specifically to include massive neutrinos and models of dynamical dark energy parameterized with $w_0w_a$ \cite{CHEVALLIER_2001, Linder_2003}. This framework uses N-body simulations to generate theoretical predictions, but the expected precision of future datasets introduces challenges regarding the accuracy and range of said simulations.

Using simulations as theoretical models has advantages, but is computationally expensive: running N-body simulations for every cosmology in a Markov-Chain Monte Carlo (MCMC) analysis, for example, would not be an efficient use of time and computational resources. In light of this, emulation of summary statistics from N-body simulations has become increasingly popular in recent years. An emulator allows one to interpolate between measurements made from relatively small sets of simulations run at a few cosmologies to obtain accurate predictions over the entire range of cosmologies spanned by the simulation suite. Emulators are fast and can be very accurate, but their error is highly dependent on the size of the dataset used to train them. To achieve an emulator error below some threshold over some range of cosmologies and wavenumbers, one must consider the cosmological parameter space volume and dimensionality.

Simulations require trade-offs between simulation volume, resolution, and run-time. The random initial conditions used, coupled to the finite box size, introduce sample variance in ensemble-averaged quantities. This problem worsens as more physics is added. Fortunately, sample variance is most significant in the regime where analytical models perform well.
Reducing the statistical error of individual simulations allows for a broader range of cosmologies to be simulated at a fixed computational cost, making emulators
more accurate. We can then assess the constraints this places on simulations: performance versus simulation size, tracer density, bias, and related parameters, which will allow us to develop a budget for simulations required to meet a given degree of accuracy over a range of cosmologies.

The outline of the paper is as follows. Section \ref{sec:LPT_HEFT} gives a brief introduction to our theoretical models, Lagrangian perturbation theory (LPT) and HEFT. Section \ref{sec:Motiv_Acc_Req} motivates our emulator accuracy requirements, including discussion of statistical errors, efficacy of second-order biasing, and errors from intrinsic alignments (IAs) and baryonic feedback. Section \ref{sec:Sims} discusses details around running N-body simulations, and gives recommendations for box size, resolution, and starting redshifts for simulations run to create HEFT emulators. Section \ref{sec:Emulator} describes the details of our emulator construction and $w_0w_a{\rm CDM}+m_{\nu}$ parameter space volumes. Section \ref{sec:Results} presents the accuracy of our emulator as a function of training set size and parameter space volume, giving concrete examples for applications to CMB lensing, galaxy-galaxy lensing, and cosmic shear.  Finally, section \ref{sec:conclusions} provides our conclusions and recommendations.  Some technical details are relegated to the appendices.

The fiducial cosmology used throughout this paper will be a $\Lambda$CDM cosmology with model parameters $\Omega_c=0.25337$, $\Omega_b=0.04933$, $h=0.68$, $\sigma_8=0.82$, $n_s=0.96$, and $\tau_{\rm reio}=0.054308$. Unless otherwise noted, distances are comoving in $h^{-1}$Mpc units, and all power spectra will be plotted at $z=0.633$. The code used for the analysis in this paper is publicly available at \href{https://github.com/AlexaBartlett/HEFT-Simulation-Budgeting}{HEFT-Simulation-Budgeting} \faicon{github}.

\section{Lagrangian perturbation theory and hybrid effective field theory}
\label{sec:LPT_HEFT}

\subsection{Lagrangian perturbation theory}

\begin{figure}
    \centering
    \resizebox{\columnwidth}{!}{\includegraphics{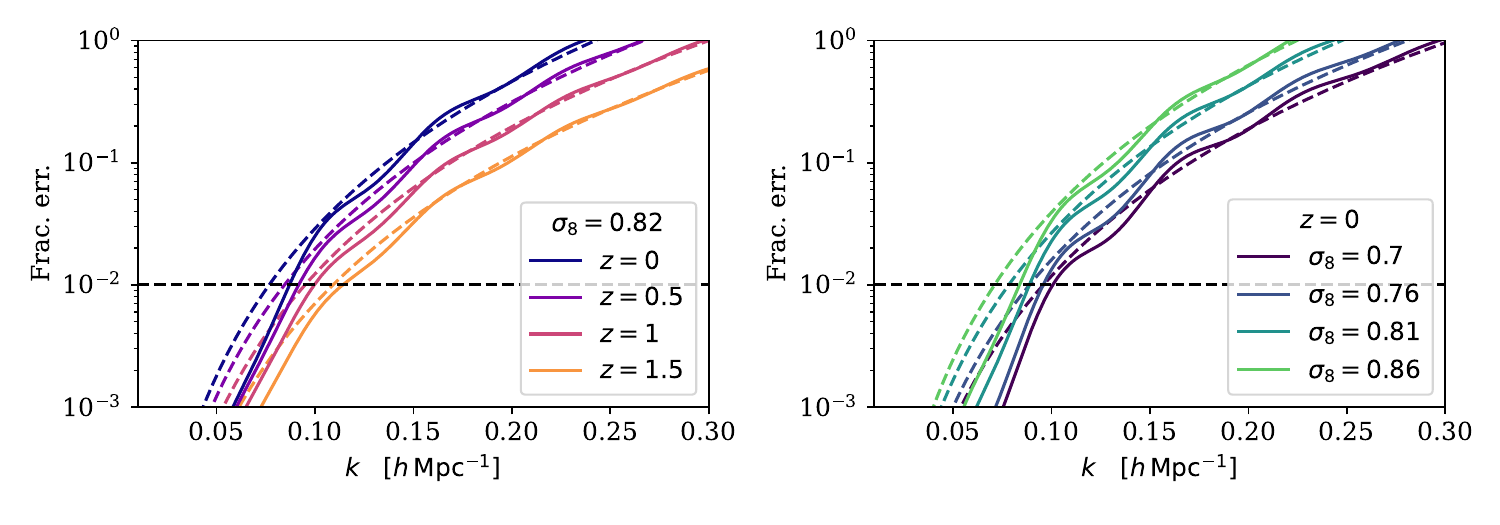}}
    \caption{Estimated theoretical error in the matter power spectrum due to neglected two-loop contributions. The dashed lines are $\propto k^4$, and the solid lines are given by $P_{\rm 2-loop}/P_{\rm tree} \approx \left[(P_{\rm 1-loop}-P_{\rm tree})/{P_{\rm tree}}\right]^2$, with one counter-term. {\it Left: } errors at four different redshifts for our fiducial cosmology. {\it Right: } errors for our fiducial cosmological parameters, but at noted values of $\sigma_8$.
    }
    \label{fig:theory_err}
\end{figure}

The Lagrangian formulation of fluid mechanics relates the current (Eulerian) position of a particle, $\mathbf{x}$, to its initial Lagrangian position, $\mathbf{q}$, through a displacement vector field, $\Psi$. Lagrangian perturbation theory attempts to perturbatively solve
\begin{equation}
    \ddot{\Psi}(q) + 2H\dot{\Psi}(q) = -\nabla \phi(q+\Psi)\,,
\end{equation}
where $H$ is the Hubble parameter, dots denote derivatives taken with respect to cosmic time, and $\phi$ is the gravitational potential. The first-order solution is known as the Zel'dovich approximation (ZA). For a review of perturbation theory, see ref.~\cite{Bernardeau_2002}. For a more detailed description of Lagrangian perturbation theory (PT), we refer the reader to refs \cite{Desjacques_2018,modernCosmo}. For a review of the effective field theory of large-scale structure, see ref.~\cite{Ivanov22}.

Higher-order displacement terms are non-trivial to compute exactly in LPT in the presence of dark energy and massive neutrinos, as they introduce additional scales and time dependence to the linear growth factor \cite{Aviles2020, Aviles2021, Senatore2017}. However, computing these terms with kernels derived for Einstein-de Sitter (EdS) cosmologies is sufficiently accurate for cosmologies close to $\Lambda$CDM \cite{Chen_2022}. We use \texttt{velocileptors}\footnote{\href{https://github.com/sfschen/velocileptors}{https://github.com/sfschen/velocileptors}} to compute LPT predictions.

While LPT has been extensively tested against simulations, it also has an internal error estimate that we expect to be reasonably accurate on large scales. The fractional matter power spectrum error due to neglecting the 2-loop contribution can be estimated by
\begin{equation}
    \frac{P_{\rm 2-loop}}{P_{\rm tree}} \approx \left[\frac{P_{\rm 1-loop}-P_{\rm tree}}{P_{\rm tree}}\right]^2
    \quad\mathrm{when}\quad
    k\ll k_{\rm NL}
\end{equation}
which we expect to be approximately proportional to $(k/k_{\rm NL})^4$ as $k\to 0$.  Figure~\ref{fig:theory_err} reveals that this fractional ``theory error'' behaves as anticipated and increases to lower $z$ and higher $\sigma_8$ precisely as expected.
This error is consistently sub-percent-level for $k \lesssim 0.05\,h\,{\rm Mpc}^{-1}$ so we incur no penalty in switching to PT for low $k$. This will simplify our task in later sections.

\subsection{Hybrid effective field theory}

Traditionally, methods for connecting simulations to observations from large galaxy surveys employ statistical models for populating galaxies in dark matter halos within large-volume cosmological N-body simulations \cite{Wechsler18}.
One of the most commonly used methods is the halo occupation distribution (HOD) formalism, used to connect matter statistics and those of biased tracers. This method faces several limitations, including but not limited to its reliance on halo-finding algorithms, which impose stringent requirements on the resolution of N-body simulations and increase computational costs \cite{Tinker_2008, Dai_2020}.  In addition, there is no well-defined criterion for when to stop adding model complexity.\footnote{Examples of choices to be made include the statistical distribution of centrals and satellites, functional form of the mean occupation, density and velocity ``profiles'', velocity bias, several types of assembly bias or conformity, aspherical halos and anisotropic velocity dispersion, etc.} Bias expansions, on the other hand, do not suffer from the same limitations. As a result, approaches to using principles of perturbative bias expansions have become increasingly common in recent years in areas where high precision is required. Hybrid effective field theory, a method first proposed in ref.~\cite{Modi_2020} and further developed in refs.~\cite{Kokron_2021,Hadzhiyska_2021,Pellejero_Iba_ez_2023,Nicola2023,shiferaw2024uncertaintiesgalaxyformationphysics, zhou2025csstcosmologicalemulatoriii}, combines the accuracy of particle displacements $\Psi$ computed in N-body simulations with the theoretical robustness of analytic, symmetries-based, Lagrangian bias expansions.

Following ref.~\cite{Modi_2020} we adopt the Lagrangian bias description.
LPT has proven a highly effective model for biased tracers in cosmologies similar to $\Lambda$CDM \cite{Chen_2021, White_2014}, and it meshes well with N-body simulations, which by their very nature provide a Lagrangian description of the matter field. In cosmologies with massive neutrinos, dark matter halos and galaxies are best modeled as biased tracers of the CDM-baryon field \cite{Castorina_2015, Bayer_2021, Desjacques_2018} rather than the total matter field. The density field of a biased tracer can be expressed directly in terms of the advected operators $\mathcal{O}_i \in \{\mathbf{1}_{cb},\,\delta,:\!\!\delta^2\!\!:,:\!\!s^2\!\!:,\nabla^2\delta,:\!\!\delta^3\!\!:,\cdots\}$ as \cite{Modi_2020}
\begin{equation}
    1 + \delta_t(\mathbf{k},a) = \sum_{\mathcal{O}_i}b_{\mathcal{O}_i}\mathcal{O}_i
    \,,
\label{eqn:bias_exp}
\end{equation}
where $b_{\mathcal{O}_i}$ are the biases associated with each operator, $\mathbf{1}_{cb}$ refers to the combined CDM and baryon density field and $:\!\!\mathcal{O}\!\!:$ refers to normal ordering, i.e.\ subtracting the expectation values ($:\!\!\delta^2\!\!:=\delta^2-\sigma^2$, $:\!\!\delta^3\!\!:=\delta^3-3\sigma^2\delta$, etc.).
The cross-spectra of two biased tracers can then be expressed as
\begin{equation}
    P_{ab}(k) = \sum_{\mathcal{O}_i,\mathcal{O}_j} b^{a}_{\mathcal{O}_i}b^b_{\mathcal{O}_j}P_{ij}(k)\,,
\end{equation}
where we have defined the basis spectra
\begin{equation}
    P_{ij}(k)(2\pi)^3 \delta_D(k+k') = \langle \mathcal{O}_i(\mathbf{k}) \mathcal{O}_j(\mathbf{k'}) \rangle\,.
\end{equation}
Notably, none of these expressions depend on how the displacements $\Psi$ have been computed. In real space for sufficiently low-bias tracers, where eq.~\ref{eqn:bias_exp} holds, it is the perturbative calculation of $\Psi$ in LPT that limits the range of scales that can be modeled. On the other hand, N-body simulations solve discretized versions of the same equations of motion that form the basis of LPT. Furthermore, all of the components that are used in the above expressions can be directly computed from N-body simulations: the displacements are simply the difference between each particle’s position and the grid point at which it began, and the linear fields used in eq.~\ref{eqn:bias_exp} can be directly computed from the initial conditions used to initialize the simulation. Thus, we can replace perturbative computations of $\Psi$ and the EdS approximation and use N-body simulations to compute the basis spectra, $P_{ij}$, an approach which has become known as HEFT \cite{Hadzhiyska_2021}. 

In principle, the HEFT bias expansion extends to arbitrary order and needs to be truncated to be practical. We retain all terms up to and including second order, plus a single cubic operator, $:\!\!\delta^3\!\!:$.  We explore the gains from including this cubic operator in section \ref{sec:HEFT_validation}.

\section{Motivation of accuracy requirements}
\label{sec:Motiv_Acc_Req}

In this section, we quickly review what is known about the anticipated uncertainties on the matter and galaxy clustering arising from measurement errors and uncertain modeling, in order to motivate a goal for the accuracy of our emulator.  If we focus on projected galaxy clustering, cosmic shear, galaxy-galaxy lensing, and lensing of the CMB, then the statistical errors come from a combination of sample variance (on large angular scales) and shot noise, shape noise or reconstruction noise (on small angular scales, for clustering, galaxy-based or CMB-based lensing, respectively).  We discuss this in the next subsection.

Uncertainties arising from our inability to model complex astrophysical processes also limit the accuracy of our cosmological inferences and thus the requirements on our emulator.  These depend upon the probe and statistic being considered, but include (1) measuring shapes or marginalizing over foregrounds, (2) redshifts, e.g., \ photometric redshifts (photo-$z$s), (3) impact of uncertain baryonic physics, (4) intrinsic alignments, and (5) non-linear modeling. The uncertainty due to the first two items can be included in our statistical error \cite{Mandelbaum18,Pratt25,Truttero_2025,SO_LAT25,zhang2025forecastingimpactsourcegalaxy}.  We will discuss each of the remaining issues in turn.  Our final error budget will motivate a goal of $\approx 1\%$ precision in the matter or galaxy auto- and cross-spectra over scales $0.05 < k < 1\,h\,\mathrm{Mpc}^{-1}$. On larger scales, we can safely switch to analytic models, while on small scales, our emulator error will be subdominant to other uncertainties.  We summarize these considerations in section \ref{sec:model_summary}.

\subsection{Statistical errors from future surveys}\label{subsec:stat_errs}

The accuracy requirements for our emulator will be largely determined by how accurately clustering, lensing, and cross-power spectra are measured by current and future surveys, as well as by other processes contributing to these signals.
On large scales and assuming Gaussian fluctuations, the statistical variance of any auto- or cross-spectrum is proportional to the square of the total power (including noise), or products of the power spectra of each component.  For example, for the galaxy-galaxy lensing (GGL) or CMB lensing--galaxy cross-power spectra the error is given by
\begin{align}
{\rm Var}\left[C^{\kappa g}_{\ell}\right] &= \frac{1}{(2\ell+1)\Delta \ell\, f_{\rm sky}}\bigg[(C_\ell^{\kappa \kappa}+N_{\ell}^{\kappa,\gamma})\left(C_{\ell}^{gg} + N^{g}_{\ell}\right) + (C_{\ell}^{\kappa g})^2\bigg]
\label{eqn:stat_err} \\
&= \frac{(C_\ell^{\kappa g})^2}{(2\ell+1)\Delta \ell\, f_{\rm sky}}
\ \frac{1+r^2}{r^2} \, ,
\end{align}
where $N^{g}_{\ell} = \bar{n}_{\rm gal}^{-1}$ is the Poisson shot noise of the galaxy sample, $N_{\ell}^{\kappa}$ is the noise power spectrum for CMB lensing, $N_\ell^\gamma = \sigma_\epsilon^2/n_{\rm eff}$ is the variance from shape noise for a sample with galaxy ellipticity dispersion $\sigma_\epsilon$ and effective source number density $n_{\rm eff}$, $f_{\rm sky}$ is the fraction of the sky over which the spectra are measured, and $\Delta \ell$ is the $\ell$-bin width.  In the second line we have rewritten the term in square brackets in terms of the cross-correlation coefficient $r=C_\ell^{\kappa g}/\sqrt{(C_\ell^{\kappa \kappa}+N_{\ell}^{\kappa,\gamma})\left(C_{\ell}^{gg} + N^{g}_{\ell}\right)}$.  For small $r$ the fractional error on $C_\ell^{\kappa g}$ scales as $r^{-1}$.  Equation \ref{eqn:stat_err} should be an underestimate of the true uncertainty, and so is conservative for our purposes.

\begin{figure}
    \centering
    \includegraphics[width=\linewidth]{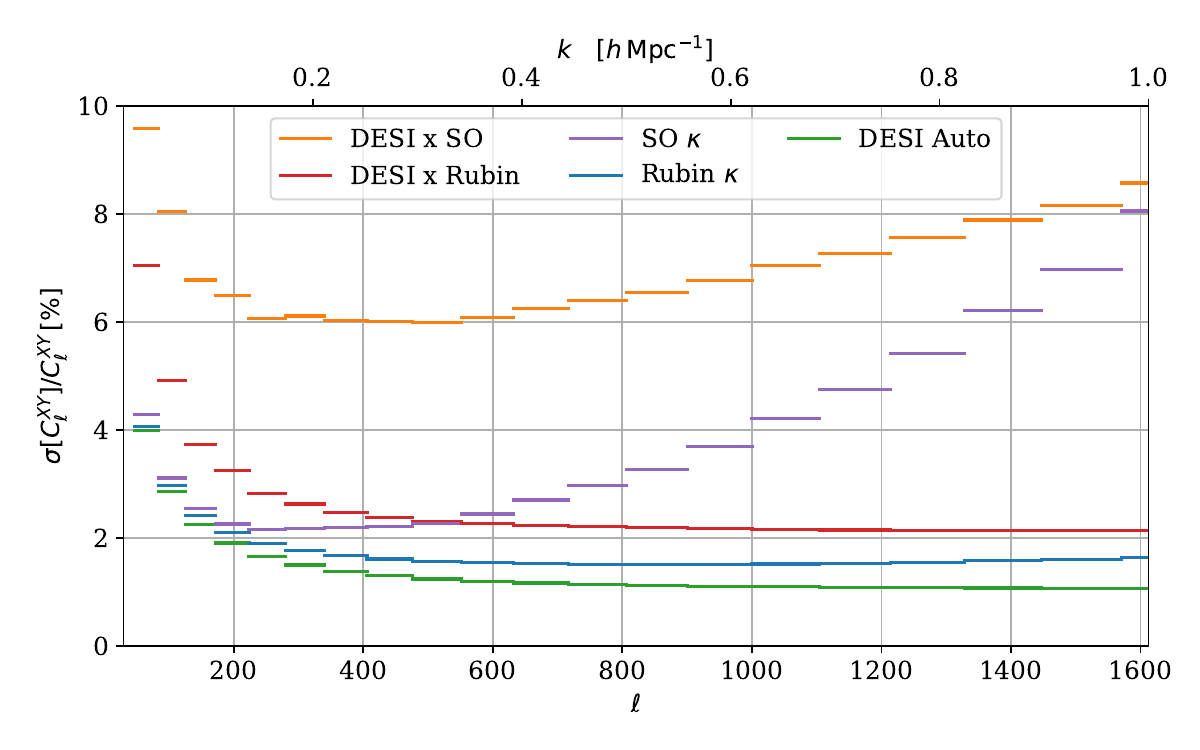}
    \caption{Forecasted fractional uncertainties on angular galaxy and CMB lensing, and galaxy clustering, auto and cross-spectra for current and future surveys in our fiducial cosmology. For DESI, we use an extended LRG-like sample with a Gaussian $dN/dz$, with $\mu_z=0.633$, $\sigma_z=0.077$, large-scale bias $b=2$ and number density $\bar{n}_{\theta}=311\,{\rm deg}^{-2}$. We additionally assume $f_{\rm sky}=0.4$. For Rubin shear, we assume $\sigma_\epsilon = 0.26$, $n_{\rm eff}=5.54\,{\rm arcmin}^{-2}$, and $f_{\rm sky}=0.35$. For our Rubin lensing source sample, we use the second-highest-redshift LSST source bin \cite{Fang_2023}. For SO, we use the publicly available lensing noise curve. We set $\Delta \ell = \sqrt{900+10\,\ell}$. A secondary $k$-axis, computed using $k = (\ell+0.5)/\chi(z=0.633)$, is provided for orientation.}
    \label{fig:frac_stat_err}
\end{figure}

We forecast the fractional uncertainties on galaxy clustering, lensing, and cross-spectra measured from one typical configuration for future surveys in figure~\ref{fig:frac_stat_err} using eq.~\ref{eqn:stat_err}.  There are several points to note from this figure.  First, for our chosen bin size, errors on the spectra are typically percent to a few percent level on the scales shown, with the galaxy and shear auto-spectra providing the most statistically significant measurements.  Later, we will argue that the shear auto-spectra will not drive our emulator requirements due to additional modeling uncertainty, which suggests galaxy auto- or cross-spectra will largely set our goals.
Also, as we shall see later, complex astrophysical processes reduce our ability to constrain cosmology from small scales, specifically for $k \gtrsim 1\,h\,{\rm Mpc}^{-1}$.

For our example forecasts, we will often use the BGS and LRG samples of DESI \cite{desicollaboration2016desiexperimentiiinstrument,desicollaboration2016desiexperimentisciencetargeting} or the extended samples described in ref.~\cite{Zhou_2023}.  In figure \ref{fig:frac_stat_err} we highlight the $z\approx 0.6$ LRG sample.  The DESI auto-spectrum for this single extended-LRG-like bin is signal-dominated, as in $C_\ell^{gg} > N_\ell^g$, to $k \approx 0.55\,h\,{\rm Mpc}^{-1}$ for this sample. However, as is evident in figure~\ref{fig:frac_stat_err}, the fractional error imparted on the auto-spectrum plateaus at $\lesssim 0.02$ for our choice of $\ell$-binning, suggesting there is little need for denser galaxy samples for cosmological constraints from galaxy clustering or cross-correlations of these galaxies with CMB lensing or cosmic shear. This raises the possibility of choosing samples that have compact and well-defined $dN/dz$, which we will see reduces other uncertainties. 

Second, the DESI $\times$ SO CMB lensing error in figure~\ref{fig:frac_stat_err} is significantly greater than the DESI $\times$ Rubin GGL error.  We have argued above that this is not driven by shot noise from the lens galaxy sample.  We are also signal-dominated to $\ell\approx 1500$ for Rubin shear with the assumed $n_{\rm eff}$.  However, CMB lensing from SO is only signal-dominated for $\ell \lesssim 350$ after which the noise grows rapidly.  Additionally, the SO $\times$ DESI CMB lensing cross-spectrum is suppressed by the relatively low $z$ and narrow nature of our galaxy distribution.   The CMB lensing kernel is very broad and peaks around $z=2$ \cite{Lewis2006, Hanson2010}, so very little of the matter contributing to CMB lensing is traced by our galaxy sample. The galaxy weak lensing kernel, however, has most of its support at much lower redshift, providing good overlap between the kernel and our galaxy $dN/dz$.  As a result, the cross-correlation coefficient $C_\ell^{\kappa g}/ \sqrt{C_\ell^{\kappa \kappa}C_\ell^{gg}}$ for GGL is 1.5--2.5 times larger than for CMB lensing for $\ell < 3000$.  This combination of factors explains the significant discrepancy in the CMB lensing and GGL fractional error.

Although CMB$\times$DESI lensing has a smaller signal-to-noise ratio (SNR) than GGL, the cumulative signal-to-noise ratio is still quite high, and the CMB lensing auto-spectrum may be used to constrain $S_8$.  With $S_8$ fixed, DESI$\times$SO CMB lensing can be used as a geometric probe (see, e.g., ref.~\cite{Heydenreich2025}).  Thus, it is still useful to consider this example in what follows.

\subsection{Intrinsic alignments}

One important consideration for our accuracy requirements is the shear power spectrum error due to intrinsic alignments (IA; \cite{Troxel_2012, Joachimi_2015, Kirk_2015, Kiessling_2015, Lamman_2024, Siegel_2025}). The observed shape of a galaxy can be decomposed into two parts, the first being the shear induced by gravitational lensing (G) and the second the change in intrinsic shape (I) due to the galaxy's local environment: $\gamma = \gamma_G + \gamma_I$. Intrinsic alignments give rise to two possible contributions, the first being intrinsic shape-intrinsic shape correlations between galaxies that are physically near one another, and the second being correlations between the intrinsic shape of one galaxy and the gravitational shear of another in a neighboring line of sight. The observed, shear-shear, E-mode angular power spectrum including IAs can then be written as\footnote{We neglect the magnification term here for simplicity, as it does not affect any of our arguments.}
\begin{equation}
    C_{ij}^{\gamma,{\rm EE}}(\ell) = C_{ij}^{\rm GG}(\ell) + C_{ij}^{\rm GI}(\ell) + C_{ij}^{\rm II}(\ell).
\end{equation}
while the cross-correlation with galaxy density contains only a GI term.  Assuming the Limber approximation, the IA $C(\ell)$s are given by
\begin{equation}
    C_{ij}^{\rm GI}(\ell) = \int_0^{\chi_H} d\chi \frac{W^i(\chi)n^j(\chi)}{\chi^2}P_{\rm GI}\left(\frac{\ell+1/2}{\chi},z(\chi)\right)
\end{equation}
and
\begin{equation}
    C_{ij}^{\rm II}(\ell) = \int_0^{\chi_H} d\chi \frac{n^i(\chi)n^j(\chi)}{\chi^2}P_{\rm II}\left(\frac{\ell+1/2}{\chi},z(\chi)\right),
\end{equation}
where
\begin{equation}
    W^i(\chi) = \frac{3H_0^2 \Omega_m}{2c^2} \chi [1+z(\chi)] \int_{\chi}^{\chi_h} d\chi ' n^i(z(\chi'))\frac{dz}{d\chi'}\frac{\chi ' - \chi}{\chi'}
\end{equation}
is the lensing kernel, and $n^i(\chi)$ is the source galaxy redshift distribution in tomographic bin $i$. 
The most widely used IA model in the literature is known as the nonlinear alignment model (NLA), which assumes that the ellipticity of a galaxy is linearly related to the gravitational potential at the time the galaxy formed. In this model, the GI and II power spectra have the same shape as the nonlinear matter power spectrum, but with a normalization $A_1(z)$ via:
\begin{equation}
    P_{\rm GI}(k,z)=A_1(z)P_{\rm \, NL}(k,z)
    \,, \quad
    P_{\rm II}(k,z)=A^2_1(z)P_{\rm \, NL}(k,z)
    \,.
    \label{eqn:NLA}
\end{equation}
The factor $A_1(z)$ is the linear response to the tidal field and is assumed to be 
\begin{equation}
    A_1(z) = -a_1 \bar{C}_1 \frac{\rho_{\rm crit,0}\Omega_{m,0}}{D(z)} \, \simeq -a_1\,\Omega_{m,0}\,\frac{0.014}{D(z)} ,
\end{equation}
where $D(z)$ is the linear growth factor, $\rho_{\rm crit,0}$ is the critical density today, $a_1$ is a free parameter of order 1, and $\bar{C}_1$ is a normalization factor conventionally set such that the dimensionless quantity $\bar{C_1}\rho_{\rm crit}\simeq 0.014$.\footnote{What we refer to as $a_1$ in this work is sometimes referred to as $A_1$ or $A_{\rm IA}$ in other works \cite{2018MNRAS.474..712M, Lamman_2024, Siegel_2025}.  We caution the reader that other conventions for $A_1(z)$ exist --- see ref.~\cite{Lamman_2024} for further details.} For reviews of intrinsic alignments, we refer the reader to the references above.

In linear theory, intrinsic alignments are exactly degenerate with our cosmological signal.  Thus, they form a potentially major obstacle to inferring cosmology from cosmic shear unless we can either limit the size of IAs by external means or break the degeneracy using scale- or redshift-dependence.  Unfortunately, the direct constraints on IAs are observationally expensive to obtain, and upper limits on their size for samples of direct interest are at best $\sigma(a_1)\sim 1$ \cite{Lamman_2024, Siegel_2025}. As in the case of galaxy bias, symmetries-based arguments can be used to define a basis of operators and bias\footnote{The models also contain additional parameters beyond just the bias parameters.} parameters (with 6 being required at quadratic order but 3 sufficing for the error bars of current data) which contribute to the correlation function of IAs \cite{Vlah2020, Chen2024}. These methods are used to handle the scale-dependence of IAs.
This leaves the redshift dependence.

Most analyses assume a rigid dependence of the IA amplitude on redshift, such as the IA parameters evolving as a power law in redshift. They then use the tight constraint on IAs from highly overlapping source-lens bin pairs to model the IA contamination over a broad redshift range. In this way, these analyses can seemingly remove IA contamination as a source of error in their analyses, even if at any given redshift the IA contribution is large. In reality, the ability to mitigate IA contamination depends sensitively on the form of the model used for the redshift evolution of the IA signal. If one instead relaxes the assumptions about redshift evolution, then IAs again contribute systematic errors at the level of the contamination at any given redshift \cite{Chen_2024}. In GGL, IA contributions can be suppressed by only using source-lens bin pairs that do not overlap significantly. However, IA contributions to cosmic shear cannot be suppressed in a generic manner, and thus set a systematic error floor of $\sim 1\%$ if one assumes $\sigma(a_1)\simeq 1$. 

The magnitude of this error is also highly dependent on both the mean redshift and the width of the source sample. The fractional $C_\ell^{GI}$ contribution decreases rapidly with an increase in mean redshift, but increases with an increase in bin width. The fractional $C_\ell^{II}$ contribution also decreases, albeit more slowly, with increasing mean redshift. However, it also decreases with increased $dN/dz$ width. For example, while the fractional shift due to IAs for the second-highest-redshift LSST source bin is $\lesssim 1\%$, it is $\gtrsim 10\%$ for the lowest-redshift bin centered around $z=0.35$ with $\sigma \simeq 0.35$.  The IA contribution can also be quite significant for shear-shear cross-spectra between $z$-bins, because the shapes of the galaxies in the lower $z$ sample are correlated with the matter lensing the higher $z$ sample.

Since IAs present such a significant contribution to the uncertainty in the power spectrum inferred from cosmic shear, we will not include cosmic shear as a driver of our power spectrum accuracy goals, though we will discuss it later.  Instead, we focus on the accuracy requirements for galaxy-galaxy and CMB lensing experiments. 
For our forecasts, we will often use parameters inspired by the BGS and LRG samples of DESI.  We have already seen that such samples are sufficiently numerous to be nearly sample variance limited on the scales of interest.  Such galaxies are also quite luminous, which reduces the impact of systematic errors from photometry of a given depth.  Further, they can have very accurate photo-$z$s, meaning one can choose compact $dN/dz$ for these samples from photometry (and the resulting $dN/dz$ can be very well calibrated spectroscopically).  An argument against the use of bright and/or red galaxies for GGL analysis is that IAs for bright, elliptical galaxies are larger than those of spirals.  Of course, what ultimately matters is the uncertainty in IA amplitude on the scales being used for cosmological inference, and it is not clear which samples will minimize this uncertainty.  Additionally, if one wishes to drop $C^{\gamma \gamma}_{\ell}$ and consider a $2\times 2$pt analysis rather than a $3\times 2$pt analysis, as we do here, IAs may be sufficiently suppressed through choosing largely disjoint $dN/dz$ such that this concern is not as relevant.  We will see later that choosing these more luminous and more biased objects is also conservative for setting our error budget from the bias expansion.

\subsection{Baryons and feedback}
\label{sec:Baryons}

\begin{figure}
    \centering
    \includegraphics[width=0.8\linewidth]{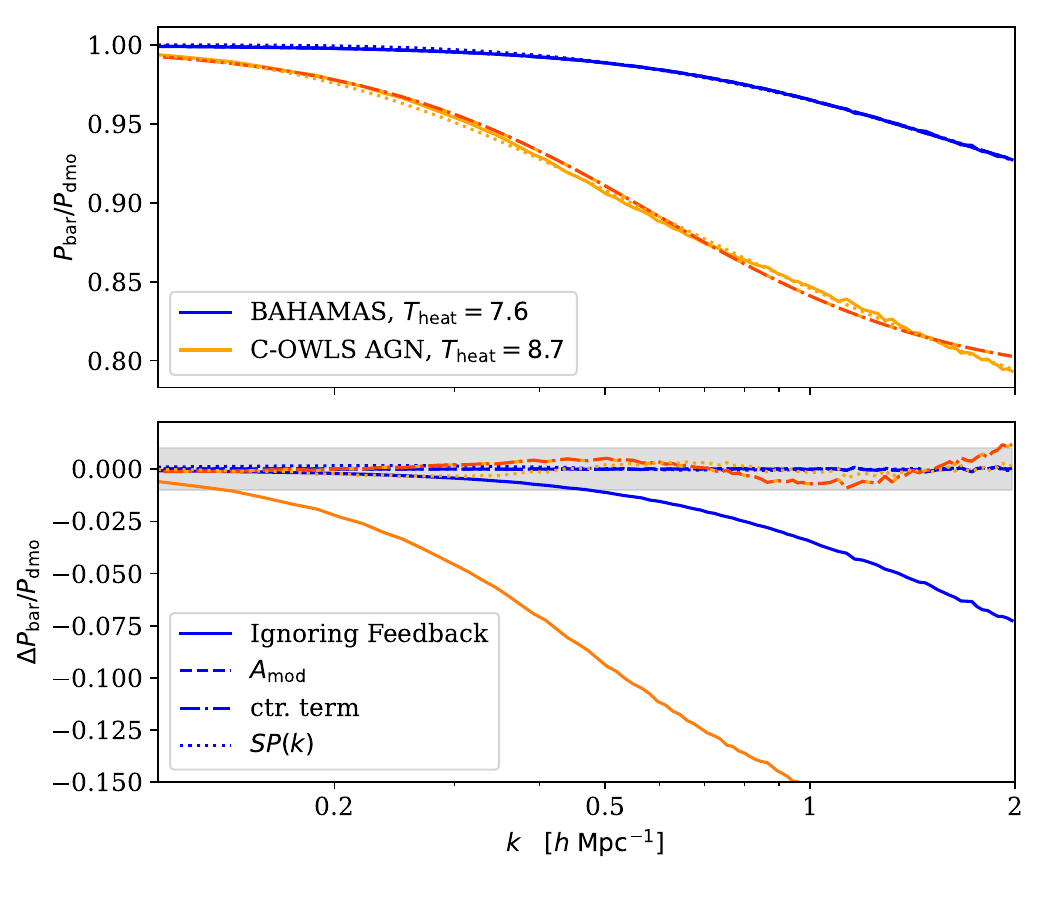}
    \caption{Comparison between the counterterm model in eq.~\ref{eqn:ctrterm}, the SP$(k)$ model, the $A_{\rm mod}$ model given by eq.~\ref{eqn:Amod}, and measurements from two hydrodynamical simulations at $z=0.25$. The top panel shows the ratios of the power spectra with feedback to those ignoring feedback. The bottom panel shows the errors arising from ignoring feedback (solid), and from modeling feedback using the $A_{\rm mod}$ (dashed), the counterterm model (dot-dashed), and the SP$(k)$ model (dotted).  The shaded grey band shows $\pm 1\%$. The blue and orange curves represent measurements from and fits to the BAHAMAS $T_{\rm heat}=7.6$ and C-OWLS with $T_{\rm heat}=8.7$ hydrodynamical simulations, respectively. The counterterm model fit to C-OWLS is plotted in a darker orange because it appears identical to the $A_{\rm mod}$ fit by eye. The measured power spectra were obtained from the \href{https://powerlib.strw.leidenuniv.nl/}{Power Spectrum Library} \cite{van_Daalen_2019} and provide a crude estimate of the current range of uncertainty in the models. We expect feedback to be stronger at lower $z$ and weaker at higher $z$.
    }
    \label{fig:feedback}
\end{figure}

One of the most significant sources of uncertainty in the power spectrum on smaller scales is the effect of baryons \cite{White_2004, Zhan_2004, van_Daalen_2011, Chisari_2018, van_Daalen_2019, Kovač2025}. Current observations \cite{preston2023nonlinearsolutions8tension,hadzhiyska2025evidencelargebaryonicfeedback,Ried-Guachalla25,efstathiou2025powerspectrumthermalsunyaevzeldovich,Pandey25, Siegel_2025_ksz, Broxterman_2025,Siegel_2025_ksz} and simulations \cite{Le_Brun_2014, McCarthy_2016, delgado2023predictingimpactfeedbackmatter, Flamingo, 2023MNRAS.524.2539P, 2025arXiv250116983B, 2014MNRAS.444.1518V} suggest that baryons begin to alter the nonlinear power spectrum at $k \sim 0.1\;h\;{\rm Mpc}^{-1}$, and the error on the power spectrum resulting from ignoring baryonic feedback may reach $\sim 1\%$ even on mildly nonlinear scales. 
This error can lead to $>5\sigma$ biases in cosmological parameters such as $S_8$ and $\Omega_m$, and to false detections of exotic physics \cite{Truttero_2025}. One could consider making stringent scale cuts in an attempt to ignore baryonic feedback entirely, but this results in a significant loss of information.
One may instead attempt to include the effects of baryonic feedback in the model for the power spectrum.  Several such models have been proposed.  For example, ref.~\cite{Chen_2024} suggested following the common approach in cosmological perturbation theory and introducing counterterms constrained by symmetry; ref.~\cite{Salcido_2023} proposed a parameterized fit to hydrodynamical simulations, which they dub the SP($k$) model; while refs~\cite{Amon_2022,schaller2025analyticredshiftindependentformulationbaryonic} suggested modeling the impact of baryons as proportional to the difference between the linear and non-linear power spectra.  We compare these models below and in figure~\ref{fig:feedback} as illustrative examples of the wide range of models currently in the literature.

The counterterm model is simply
\begin{equation}
    P_m(k,z) = P_{\rm \, NL}(k,z)\left(1-\frac{b_{\nabla^2m}\,k^2}{1+(kR)^2}\right)\,.
\label{eqn:ctrterm}
\end{equation}
with two free parameters, $b_{\nabla^2m}$ and $R$, which physically correspond to a sound speed and the characteristic length scale over which baryonic effects alter $P_m$.  Since our bias model already includes $b_{\nabla^2}$, which is degenerate with $b_{\nabla^2m}$, there is only one new parameter overall.  Ref.~\cite{Chen_2024} showed that this simple model allows one to fit a broad range of feedback scenarios, a result we confirm in figure~\ref{fig:feedback}.

The equations for the SP($k$) model are much more complex, and can be found in ref.~\cite{Salcido_2023}.  We will see below that empirically it can be well fit by $P_m/P_{\rm dmo}\simeq 1 - \mathrm{(const)}k^2 + \cdots$ though due to the functional forms chosen there are many non-analytic functions (e.g.\ exponentials of $\lambda\log_{10}k$ which transform into $k^{\lambda/\ln(10)}$) which makes series expansion difficult.

For the $A_{\rm mod}$ model, ref.~\cite{Amon_2022} initially decomposed the total matter power spectrum into 
\begin{equation}
    P_m(k,z) = P_{\rm \, L}(k,z) + A_{\rm mod}\left[P_{\rm \, NL}(k,z)-P_{\rm \, L}(k,z)\right] \,,
\label{eqn:PNL_baryons}
\end{equation}
with $P_{\rm \, L}(k,z)$ and $P_{\rm \, NL}(k,z)$ the linear and nonlinear matter power spectra, respectively, and $A_{\rm mod}$ a constant independent of cosmology or redshift.  Ref.~\cite{schaller2025analyticredshiftindependentformulationbaryonic} find that promoting the constant $A_{\rm mod}$ to
\begin{align}
    A_{\rm mod}(k) &= 0.745 + 0.1275\left[1-{\rm tanh}\left(\frac{{\rm log}_{10}(k/k_{\rm mid})}{0.656}\right)\right]
    \label{eqn:Amod} \\
    &\simeq 0.745 + 0.1275\left[ 1-\tanh\ln\left\{\frac{k}{k_{\rm mid}}\right\}^{0.662} \right] \\
    &= \frac{1+0.745\,(k/k_{\rm mid})^{1.324}}{1+(k/k_{\rm mid})^{1.324}}
    \simeq 1 - 0.255\left(\frac{k}{k_{\rm mid}}\right)^{1.324} + \cdots
\end{align}
provides a good fit to the impact of baryons on a range of {\tt Flamingo} simulations \cite{Flamingo}. The free parameter $k_{\rm mid}$ may be tuned to optimally fit different individual simulations.  We shall consider the original variant, $A_{\rm mod}=$ constant, below, since we won't be using the {\tt Flamingo} simulations, but we note that in general $A_{\rm mod}$ may need to be $k$- and $z$-dependent.

All of these models ``predict'' that the impact of baryons at low $k$ must be proportional to $k^2\,P(k)$, essentially due to the Jeans argument.\footnote{To characterize Jeans' argument crudely, pressure forces scale as $p\sim c_s^2\rho\sim c_s^2\nabla^2\phi$ so gradients of pressure are suppressed, compared to gradients of gravitational potential, by $k^2$.  To the extent the simulations correctly incorporate this physics and the fits explain those data, we expect a $k^2$ correction as in Eq.~\ref{eqn:baryon_taylor}.  Many fitting functions do not enforce this explicitly, though in order to provide a good fit to simulations, they do implicitly.}  This is explicit in the counterterm model.  To see this for the $A_{\rm mod}$ model, note that from perturbation theory we expect $P_{NL}-P_L\simeq c_{\rm nl}k^2P_L$ in the quasi-linear regime for some constant $c_{\rm nl}$. Thus, eqs.\ \ref{eqn:ctrterm} and \ref{eqn:PNL_baryons} may both be written as
\begin{equation}
  P_m(k,z) \simeq P_{\rm \, NL}(k,z)\left[ 1-c_2^{(b)}\left(kR_b\right)^2 + c_4^{(b)}\left(kR_b\right)^4 + \cdots\right]
  \,, \quad k\ll R_b^{-1},\ k_{NL}\,,
\label{eqn:baryon_taylor}
\end{equation}
where $c_2^{(b)}$ is a constant of order unity if $R_b$ is the characteristic scale of baryonic feedback ($R_b\sim 1\,h^{-1}$Mpc).  On general grounds, the corrections should form a power series\footnote{This can be resummed into a Pade approximant, as in Eq.~\ref{eqn:ctrterm}, if desired.} in even powers of $kR_b$, whose coefficients should be $\mathcal{O}(1)$ if $R_b$ is appropriately chosen.  Detailed models of baryonic effects can then be used to determine $R_b$.

The three baryonic feedback models discussed in this subsection are compared to measurements from two hydrodynamical simulations in figure~\ref{fig:feedback}. We have chosen these two particular hydrodynamical simulations for illustration, since they represent simulations near the ``low'' and ``high'' end of the predictions from the currently available runs (though see below for caveats).  Specifically, we use the BAHAMAS $T_{\rm heat}=7.6$ \cite{McCarthy_2016} model as an example of a simulation predicting a ``low'' feedback impact on $P_m$ and the C-OWLS $T_{\rm heat}=8.7$ \cite{Le_Brun_2014} model for a ``high'' impact example.

Figure~\ref{fig:feedback} shows that ignoring feedback altogether introduces at least 1\% error in the power spectrum for $k \gtrsim 0.1 \,h\,{\rm Mpc}^{-1}$ and $k\gtrsim 0.5 \,h\,{\rm Mpc}^{-1}$ for the ``high'' (orange) and ``low'' (blue) simulations, respectively. This error consistently reaches $1\%$ by $k \approx 1\,h\,{\rm Mpc}^{-1}$ even in weak feedback models.  However, with constants and counterterms appropriately set, the counterterm and $A_{\rm mod}$ models differ by $\lesssim 1\%$ to $k \lesssim 1\;h\,{\rm Mpc}^{-1}$. 
This agreement is expected, given the arguments above and that both models are known to provide good fits to hydrodynamical simulations.  The models start to differ by more than 1\% above $k=1\;h\,{\rm Mpc}^{-1}$, which suggests that emulating the power spectrum without feedback at the percent level for $k \gtrsim 1\,h\,{\rm Mpc}^{-1}$ is unnecessary since the systematic error in the feedback model will dominate over emulator error.  Below, we will see that the need to marginalize over free parameters in the model leads to the same conclusion.

Though we chose to compare the BAHAMAS $T_{\rm heat}=7.6$ \cite{McCarthy_2016} and C-OWLS $T_{\rm heat}=8.7$ \cite{Le_Brun_2014} hydrodynamical simulations in figure~\ref{fig:feedback}, there is growing evidence for more gas redistribution than the majority of the current generation of simulations predict \cite{preston2023nonlinearsolutions8tension,hadzhiyska2025evidencelargebaryonicfeedback,Ried-Guachalla25,efstathiou2025powerspectrumthermalsunyaevzeldovich,Pandey25,Eckert2025, 2025arXiv251204209B, 2025arXiv251202954S, kSZ_benchmark_feedback}. However, simply increasing the strength of feedback in existing models leads to simulated galaxies that differ in other key properties from observations, suggesting that the \emph{form} of the feedback in our current models is likely incomplete.  We thus regard the precise shape of these curves as indicative at best.

Assuming that these models, though not correct in detail, are reliable enough to estimate the characteristic scale of baryonic feedback, we can fit $P_m/P_{\rm dmo}$ to eq.~\ref{eqn:baryon_taylor} to determine $R_b$ and the $c_{2n}^{(b)}$. For example, for the $z=0.25$ output of the C-OWLS $T_{\rm heat}=8.7$ simulation shown in figure~\ref{fig:feedback}, which is close to the level of feedback consistent with current observations \cite{preston2023nonlinearsolutions8tension,hadzhiyska2025evidencelargebaryonicfeedback,Ried-Guachalla25,efstathiou2025powerspectrumthermalsunyaevzeldovich,Pandey25}, we have $b_{\nabla^2m}\simeq 0.7\,h^{-2}\mathrm{Mpc}^2$ and $R\simeq 1.8\,h^{-1}$Mpc. For eq.~\ref{eqn:baryon_taylor} we could take, for example, $R_b=1.25\,h^{-1}$Mpc, $c_2^{(b)}=0.43$ and $c_4^{(b)}=0.88$ as describing the impact of baryons when $k< R_b^{-1}$.

Marginalizing over $c_2^{(b)}$ will make the ``derived'' DMO power spectrum at $k<R_b^{-1}$ uncertain by $\sigma_{c_2}(kR_b)^2$, and similarly for $c_4^{(b)}$.  Unless we have very tight priors on $c_{2n}$, the error on $P_m$ will grow rapidly to higher $k$.  Once $ k\,R_b\simeq 1$, many terms will enter with comparable sizes, allowing almost any shape and rendering those scales unusable for precision cosmology (in the absence of external information) \cite{Garcia-Garcia24, Bigwood24, Anbajagane2025}.  For this reason, we will confine our attention to $k\lesssim 1\,h\,\mathrm{Mpc}^{-1}$.

An alternative approach is to try to simultaneously model multiple observations (e.g.\ ref.~\cite{wayland2025calibratingbaryoniceffectscosmic} for a recent example). This allows us to break model or parameter degeneracies in any given observation using ``new'' information from complementary observations. However, it risks introducing bias if the models cannot simultaneously provide high-fidelity fits to all the datasets.  This remains an open research problem.  A reasonable compromise is to use the more flexible functional forms, but impose informative priors on the coefficients based on external data.  This reduces the inflation of the error bars, but allows sufficiently powerful data to overcome the priors if they turn out to be incorrect, thus mitigating the bias (since the parameterization is capable of fitting any physically-allowed feedback).  It also allows a quantitative investigation of prior sensitivity.

\subsection{Beyond second-order biasing}
\label{sec:HEFT_validation}

\begin{figure}
    \centering
   \includegraphics[width=\linewidth]{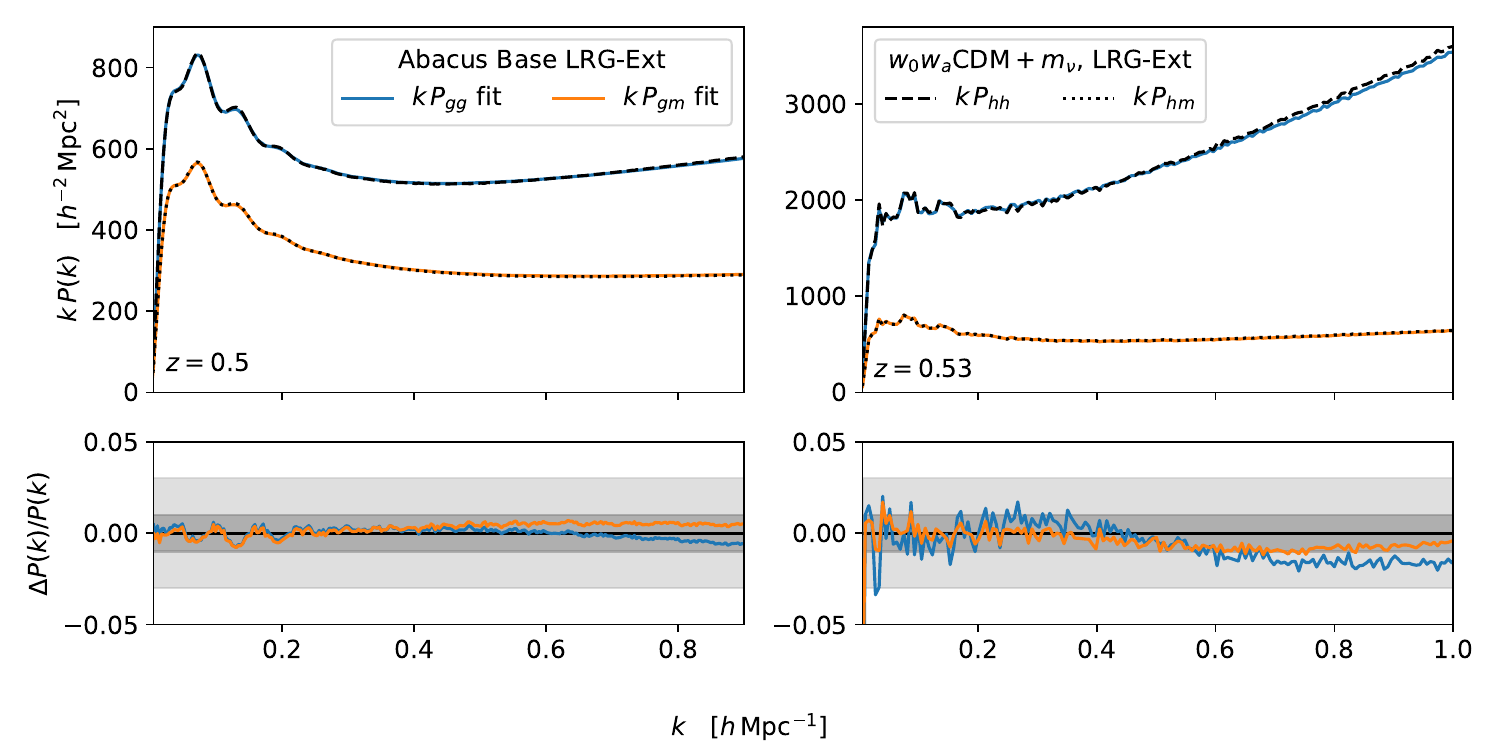}
    \caption{Plots showing HEFT (with quadratic bias) fits to spectra of mock galaxy catalogs with clustering similar to DESI LRGs in both $\Lambda$CDM and $w_0w_a\mathrm{CDM}+m_\nu$ cosmologies.  The lower panels show the fractional errors with 1\% and 3\% bands shaded in grey. The left panels show the average of the ZCV variance-reduced galaxy auto- and cross-spectra measured from 13 of the base {\tt AbacusSummit} \cite{Maksimova_2021, Garrison_2021} simulations (which assume $\Lambda$CDM). The right panel shows the HEFT fits to mock galaxy auto- and cross-spectra measured from a single $w_0w_a{\rm CDM}+m_{\nu}$ simulation.  These mock data are noticeably noisier than their $\Lambda$CDM counterparts due to the use of a single, smaller box. The best-fit biases for the Abacus simulations and $w_0w_a$ simulation are $b_{1}=1.4,\,b_2=2.6\times 10^{-2},
    \,b_s=-0.27,\,b_{\nabla^2}=-2.8\times 10^{-2},\,{\rm sn}=73$ and $b_{1}=1.76,\,b_2=2.2,
    \,b_s=-1.44,\,b_{\nabla^2}=-0.45,\,{\rm sn}=212$,
    respectively, where sn is the shot noise.
    }
    \label{fig:HEFT}
\end{figure}

HEFT is a systematic expansion; its accuracy over a given range is determined by the bias of the sample and the order of bias expansion used. As discussed in the previous subsections, the presence of systematics such as intrinsic alignments and baryonic feedback prevents us from measuring the matter power spectrum to $1\%$ accuracy beyond $k \approx 1h\,{\rm Mpc}^{-1}.$ In this subsection, we determine the appropriate order bias expansion to use given our accuracy requirements.

HEFT is more accurate at a given scale for tracers with lower and less scale-dependent bias, and as we include more terms in the bias expansion.  To provide a stringent test of the model, we will deliberately choose biases that are higher than we know could be achieved with upcoming, deep surveys.  To investigate the convergence with order of the bias expansion, we shall consider linear, quadratic, and part of the cubic order expansions.  Linear bias corresponds closely to the often used ``linear bias times power spectrum-fit'' approach, except that we use the full N-body matter power spectrum rather than a halo-model-inspired approximation to it.  Quadratic bias is the normal setting for HEFT.  Inclusion of a subset of the $3^{\rm rd}$ order terms allows us to see where those terms become relevant, and slightly improves the fit over just including the quadratic terms.

Ref.~\cite{Kokron_2021} showed that HEFT with second-order biasing is accurate to $k_{\rm max} \approx 0.6\,h\,\mathrm{Mpc}^{-1}$ for low-bias tracers in $w{\rm CDM}$ cosmologies, a finding that has been confirmed by several other works \cite{Nicola2023, zhou2025csstcosmologicalemulatoriii, DeRose_2023, Zennaro_2022, Zennaro_2023, Pellejero_Iba_ez_2023}.  We repeat this test for a mock galaxy catalog with $b\approx 2.4$ from a simulation of $\Lambda$CDM at $z\simeq 0.5$, and extend the test to auto- and cross-spectra measured from a $w_0w_a\mathrm{CDM}+m_\nu$ simulation.   The parameters for the $w_0w_a\mathrm{CDM}+m_\nu$ simulation are taken to be close to the best fit to combination of DESI+CMB \cite{DESI-DR2}, with $A_s=2.09\times 10^{-9}$, $n_s=0.97$, $\omega_{c}=0.12$, $\omega_b=0.022$, $h=0.63$, $w_0=-0.45$, $w_a=-1.52$ and three equal-mass neutrinos with $\sum m_\nu=0.059$ eV.  Including linear and quadratic bias terms, figure~\ref{fig:HEFT} shows that HEFT is capable of fitting simulation output to high accuracy up to $k_{\rm max} \approx 0.6 \,h\,\mathrm{Mpc}^{-1}$ even when the evolution in the dark energy equation of state is particularly rapid.

\begin{figure}
    \centering
   \includegraphics[width=\linewidth]{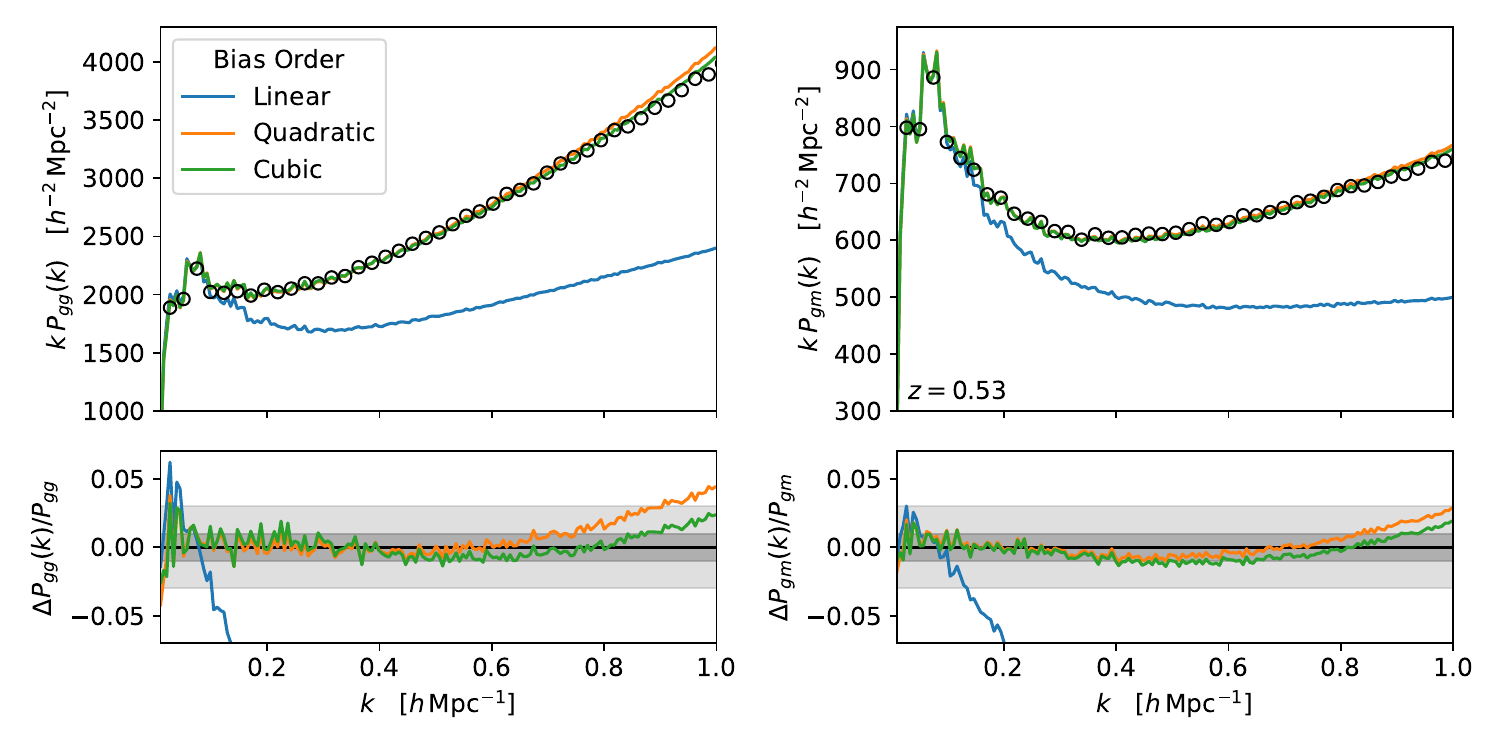}
    \caption{Plots showing the gains from including successively higher order bias terms in HEFT fitted simultaneously to $P_{gg}(k)$ and $P_{gm}(k)$ for a single $\Lambda$CDM simulation at $z=0.53$.  We choose $k_{\rm max}=0.6\,h\,\mathrm{Mpc}^{-1}$ for the quadratic and cubic models and $k_{\rm max}=0.15\,h\,\mathrm{Mpc}^{-1}$ for the linear model. Higher $k$ points are included in the fit, but are weighted such that $k\leq k_{\rm max}$ points are prioritized. The top left and right panels show the measured and fit galaxy auto- and cross-spectra, respectively, while the bottom panels showcase the fractional error in the HEFT fits.  The impressive gain in accuracy in going from linear to quadratic bias is evident, particularly in the bottom panels. The bottom panels also showcase noticeable improvements in the galaxy auto- and cross-spectrum fits at the highest $k$ when including the cubic operator $:\!\!\delta^3\!\!:$.
    The light and dark grey shaded bands in the lower panels indicate $3\%$ and $1\%$ error, respectively.
    } 
    \label{fig:CubicGains}
\end{figure}

Figure \ref{fig:CubicGains} investigates the convergence with increasing order in the bias expansion.  For this test, we use the same $\Lambda$CDM cosmology as above, also at $z\simeq 0.5$.  The blue lines show the best fits, to $k_{\rm max}=0.1\,h\,\mathrm{Mpc}^{-1}$, assuming linear bias.  This is a configuration often used in analyses of cosmic shear. However, we see that it is not particularly accurate for this sample and cannot provide a good fit to $P_{gg}$ and $P_{gm}$ simultaneously once the two fields begin to decorrelate \cite{Modi17}.  Inclusion of the quadratic bias operators provides a dramatic improvement in the fit. Furthermore, ref.~\cite{Porredon2021} found approximately $1\sigma$ shifts in their inferred value of $S_8$ when switching between a linear and quadratic bias model, and ref.~\cite{Chen_2024} found similarly sized shifts when fitting to simulations to $k=0.3\, h\, \rm Mpc^{-1}$. Increasing $k_{\rm max}$ to $0.6\,h\,\mathrm{Mpc}^{-1}$ we find that $P_{gg}$ and $P_{gm}$ can be simultaneously fit to 1\% up to $k_{\rm max}$. 

Finally, we also examine the gains from inclusion of the cubic operator $:\!\!\delta^3\!\!:=\delta^3-3\sigma^2\,\delta$, as done in refs.~\cite{zhou2025csstcosmologicalemulatoriii, shiferaw2024uncertaintiesgalaxyformationphysics}. We subtract $3\sigma^2\,\delta$ to avoid changing the linear bias, since $\delta^3$ itself contains a piece degenerate with $\delta$. In principle, multiple cubic terms are allowed by symmetry. However, this local term captures the leading-order behavior \cite{zhou2025csstcosmologicalemulatoriii, shiferaw2024uncertaintiesgalaxyformationphysics} and is straightforward to implement. We show the improvements in fit for a single $\Lambda$CDM simulation in figure~\ref{fig:CubicGains}.
The bottom panels show modest gains from inclusion of the cubic term $:\!\!\delta^3\!\!:$. For the galaxy cross-spectrum at high-$k$, the error in the cubic fit is $\lesssim 2\%$ out to $k = 1\,h\,{\rm Mpc}^{-1}$, while the error in the quadratic fit reaches $2\%$ around $k = 0.8\,h\,{\rm Mpc}^{-1}$, and is $\sim 3\%$ at $k = 1\,h\,{\rm Mpc}^{-1}$. We note, however, that the choice to include only this single cubic operator is not theoretically consistent within perturbation theory beyond one-loop order. We have verified that the gains for a single $w_0w_a{\rm CDM}+m_{\nu}$ simulation are similar.

The $k$-range over which a given bias order is accurate can be used to infer what bias order is required for a given galaxy and lensing signal-to-noise ratio and hence angular number density. For example, HEFT with linear bias provides $\sim 1\%$ accuracy to $k \simeq 0.1\,h\,{\rm Mpc}^{-1}$.  At $z\simeq 0.6$ this corresponds to $\ell\simeq 240$.  The angular galaxy auto-spectrum for our fiducial sample with $\mu_z=0.633$ and $\sigma_z = 0.077$ is equal to the Poisson shot noise at $\ell=240$ for angular number density $n_{\theta}\simeq 60\,{\rm deg}^{-2}$.  For samples less dense than around this value, the use of linear bias is likely appropriate given the error bars on the observation.  Alternatively, if one is limited to linear bias models for some reason, improving the angular density past this value provides diminishing returns.  By contrast HEFT with quadratic bias is $\sim 1\%$ accurate to $k \simeq 0.6\,h\,{\rm Mpc}^{-1}$.  At $z\simeq 0.6$ this corresponds to $\ell\simeq 1000$.  For the same sample, $C_{\ell}^{gg}$ equals the Poisson shot noise at $\ell=1000$ for $n_{\theta}\simeq 685\,{\rm deg}^{-2}$.  Depending upon the value of $f_{\rm sky}$, samples sparser than this should be well modeled by truncating the bias expansion at quadratic order.  We note that number densities sufficient for the auto-spectra will additionally be sufficient for the cross-spectra, as evident from eq.~\ref{eqn:stat_err}.

\begin{figure}
    \centering
    \includegraphics[width=\linewidth]{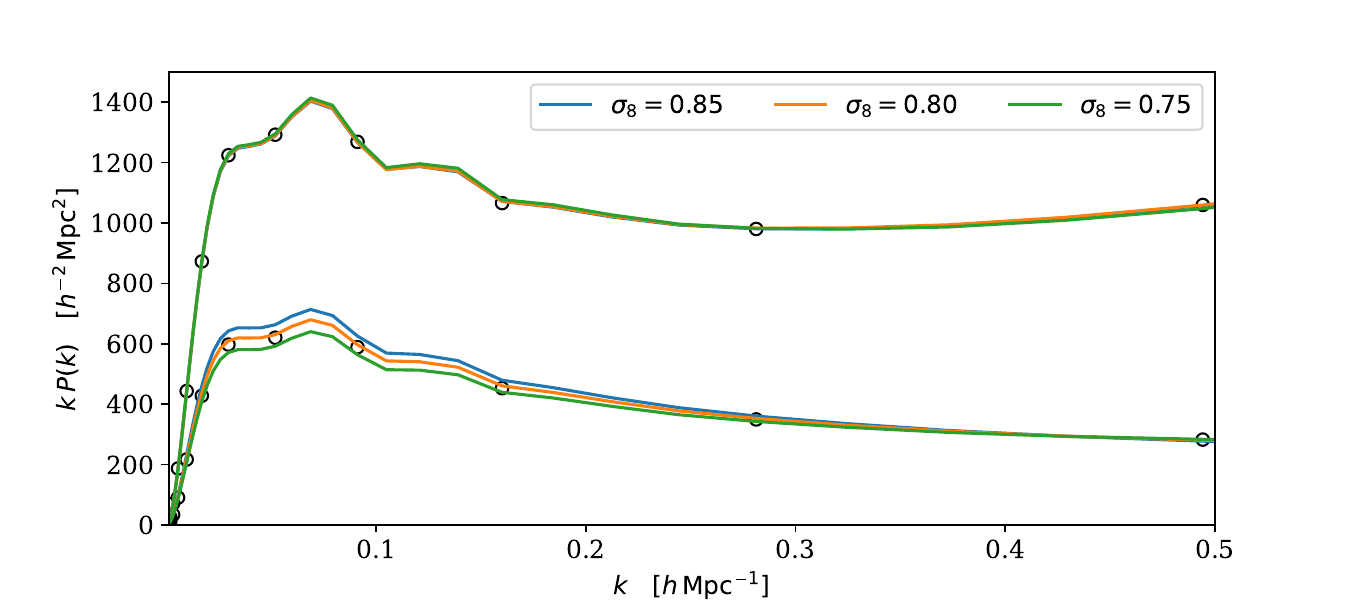}
    \caption{Best-fit $\Lambda$CDM-HEFT models to synthetic $P_{gg}$ and $P_{gm}$ data (points) for our fiducial cosmology and $dN/dz$. For each color curve, $\sigma_8$ is set to the value indicated in the legend with other cosmological parameters fixed, and the bias parameters are varied to find the best-fit $P_{gg}$ and $P_{gm}$. The models vary most significantly in their low $k$ $P_{gm}$ amplitude predictions, while making nearly identical predictions at higher $k$ due to scale-dependent bias. With such models, then, the constraining power in $\sigma_8$ comes primarily from lower $k$ than might otherwise be expected (see ref.~\cite{Chen_2022} for a similar situation with 3D clustering). }
    \label{fig:sig_8_fix}
\end{figure}

The need to include scale-dependent bias through higher-order bias terms in the model has implications for which scales contribute most to the cosmological constraints from a given dataset.  This is because the model has the flexibility to fit a range of different small-scale data, even at fixed cosmology, through modifications to the non-linear bias.  This is illustrated in Figure~\ref{fig:sig_8_fix}, which is inspired by figure~4 of ref.~\cite{Chen_2022}. In figure~\ref{fig:sig_8_fix} we have generated some synthetic data using our HEFT model, with parameters roughly consistent with our fiducial LRG sample.  Then, for several different cosmologies that differ only in the value of $\sigma_8$, we fit the model to these synthetic data at fixed cosmology (i.e.\ varying\footnote{Since these are synthetic data, we use a single set of bias parameters for both the auto- and cross-spectrum.  For observational data, one should allow $b_{\nabla}$ to differ slightly between auto- and cross-spectra as discussed in section \ref{sec:Baryons} and in section 4.2 of ref.~\cite{Sailer_2025}.  This has no impact on our conclusions.} only the ``nuisance'' parameters).  At low $k$ the model does not have enough freedom to fit the synthetic data if the value of $\sigma_8$ does not match that used to generate the data.  However, at higher $k$ there is sufficient flexibility in the model to accommodate both $P_{gg}$ and $P_{gm}$, even though the cosmology is incorrect.  Phrased another way, these data are consistent with a range of cosmologies given our current uncertainties around galaxy formation.  For this reason, the inclusion of smaller-scale data serves primarily to fix the nuisance terms rather than improve the cosmological fit.  This is a familiar behavior that has also been seen in 3D clustering \cite{Hand17,Chen_2022} and in cosmic shear analyses \cite{Anbajagane2025}.  For a concrete example from galaxy-galaxy lensing, note that the cosmological constraints from a linear-theory-only analysis restricted to large scales are almost as tight as those from a more complex model fit to a wider range of scales \cite{Sailer_2025}.

\subsection{Summary}
\label{sec:model_summary}

\begin{figure}
    \centering
    \includegraphics[width=\linewidth]{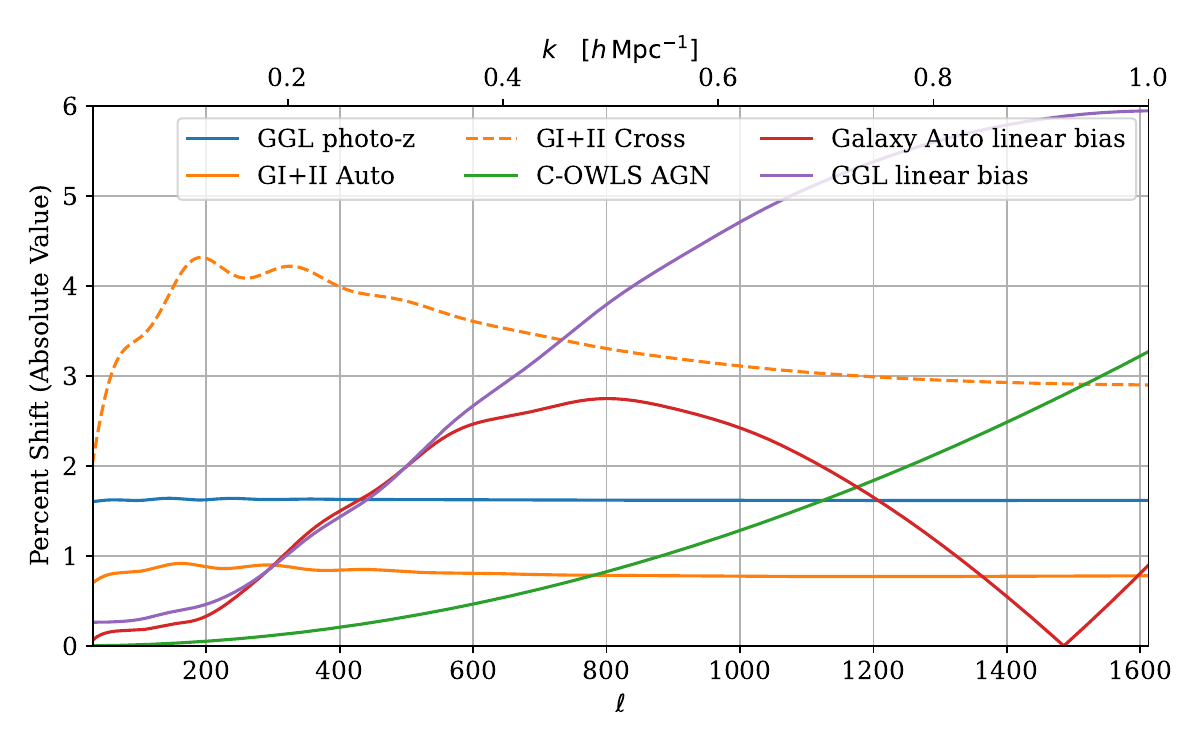}
    \caption{Fractional shifts in angular power spectra due to photo-$z$ errors (blue), intrinsic alignments (orange), incorrect baryonic feedback modeling (green), and use of linear bias rather than quadratic bias for $C_\ell^{gg}$ and $C_\ell^{
    \kappa g}$ (red and purple, respectively). Details of the analysis choices are described in the text of section \ref{sec:model_summary}.} 
    \label{fig:frac_shift}
\end{figure}

Figure \ref{fig:frac_shift} provides a summary view of the various ``modeling systematics'' discussed previously.  Since the errors are not statistical, rather than showing $1\,\sigma$ error bars, we indicate the amount by which the signal shifts if we modify the model by an amount consistent with current uncertainties (which, of course, are themselves somewhat uncertain). 

For the GGL photo-$z$ curve, we assume a $0.015$ shift in the redshift distribution of the LSST source galaxies, consistent with \cite{Giannini2025, Myles_2021, ChoppindeJanvry_2025}, and compute the resulting percent shift in $C_{\ell}^{\kappa g}$. For this photo-$z$ error, we see a $\gtrsim1.5\%$ shift in $C_{\ell}^{\kappa g}$ on all scales. The orange curves correspond to fractional shifts due to intrinsic alignments $|C_{\ell}^{GI}+C_{\ell}^{II}|/C_{\ell}^{GG}$. The solid curve shows the shift in the auto-spectrum of the second-highest-$z$ LSST source bin, while the dashed curve corresponds to the shift in the cross-spectrum between the highest-$z$ and second-highest-$z$ LSST source bins. As is evident in figure~\ref{fig:frac_shift}, this shift is significantly higher for the cross-spectrum since the galaxies in the lower $z$ bin are more correlated with the structure lensing the galaxies in the higher $z$ bin. We use the NLA with $ a_1 = 1$. 

The C-OWLS AGN curve shows the resulting shift in $C_{\ell}^{gg}$ for our LRG-ext sample if one assumes a $10\%$ error in the best-fit values of $c_2^{(b)}$ and $c_4^{(b)}$ to the C-OWLS $T_{\rm heat}=8.7$ hydrodynamical simulation, as described in section \ref{sec:Baryons}. The resulting shift in the galaxy auto-spectrum peaks just above $3\%$ at high-$k$. 

Lastly, the galaxy auto linear bias and GGL linear bias curves correspond to the shifts in $C_{\ell}^{gg}$ and $C_{\ell}^{\kappa g}$, respectively, when performing HEFT fits using quadratic versus linear bias. We calculate the best-fit linear and quadratic biases to $P_{gg}(k)$ and $P_{gm}(k)$ for our LRG-ext sample, and use the resulting combined spectra to calculate the corresponding angular power spectra. The resulting shifts in $C_{\ell}^{gg}$ and $C_{\ell}^{\kappa g}$ peak at approximately $3\%$ and $6\%$, respectively. 

Based on the considerations above, we have chosen to set a goal of $1\%$ accuracy on the HEFT spectra for $0<z<2$ and $0.05<k<1\,h\,\mathrm{Mpc}^{-1}$.  This is quite conservative and would make this error subdominant to the combined statistical, systematics, and modeling-induced errors of all next-generation surveys.  At higher redshift or smaller $k$, we can safely use 1-loop perturbation theory.  For the smaller scales and lower redshifts, we will need to use simulations and would like these simulations to achieve this accuracy for each cosmology in our training set, ensuring that any emulation or interpolation maintains this accuracy over the full parameter space.
We regard the upper limit of the $k$ range as a ``soft requirement'', since it is debatable whether HEFT is, or needs to be, $1\%$ accurate to $k=1\,h\,\mathrm{Mpc}^{-1}$.  We have chosen to be conservative in this regard since, in many cases, it does not noticeably increase the ``cost'' of running the simulations or training the emulator.  Where the costs are a function of $k_{\rm max}$, we shall illustrate the dependence explicitly.

We now turn to the flow-down of these goals to requirements on simulations in order to achieve this level of accuracy in each run (and thus the ``cost'' per simulation) and then to the number of simulations required to train an emulator capable of reproducing the HEFT spectra across the full parameter space to this accuracy.

\section{Numerical error considerations}
\label{sec:Sims}

In the previous section, we established that pure perturbation theory is not sufficiently accurate given our power spectrum error goals; therefore, we need to use a hybrid of theory and simulations as our model. We thus require that individual simulations satisfy the accuracy requirements outlined in the previous section for any set of cosmological parameters.  The accuracy and cost of individual simulations must be considered in tandem to develop a budget. Several factors influence both simulation accuracy and cost, and we will take them in turn.  Section \ref{sec:nbody_summary} summarizes our recommendations.

\subsection{Simulation volume}

Simulations must be run in a finite volume, with a particular realization of the initial conditions, which introduces sample variance in ensemble-averaged quantities such as power spectra due to the finite number of long-wavelength Fourier modes simulated.  In the sample-variance dominated limit, within linear theory,
\begin{align}
    \frac{\Delta P_m}{P_m} &=
    \sqrt{\frac{2}{N_{\rm modes}}} \qquad\qquad (k\ll k_{\rm \, NL}) \nonumber \\
    &\simeq 0.01
    \left(\frac{\Delta k/k}{0.2}\right)^{-1/2}
    \left(\frac{k}{0.1\,h\,\mathrm{Mpc}^{-1}}\right)^{-3/2}
    \left(\frac{L_{\rm box}}{1\,h^{-1}\mathrm{Gpc}}\right)^{-3/2}\,,
\label{eqn:dPonP}
\end{align}
where $k_{\rm \, NL}$ is the non-linear scale and we have assumed the simulation is a periodic cube of side length $L_{\rm box}$.  Note that $\Delta P_m/P_m$ can become large at low $k$, requiring large simulation volumes (and hence increased simulation cost if we hold mass and force resolution constant) even if we switch to perturbation theory for the very largest scales.

We can reduce the realization-to-realization scatter at low $k$ by using a combination of theoretical modeling and statistical techniques.  In particular, control variates allow for the reduction of the variance of a random variable given a correlated random variable with a known mean. 
In cosmology, it was first used to reduce the variance of measurements from $N$-body simulations, the control variate being measurements from approximate simulations \cite{Chartier_2021, Chartier_2021_2, Chartier_2022, Ding_2022}. Similar variance reduction is possible at a significantly reduced cost if one uses the Zeldovich approximation (ZA) as a control variate. This method is known as Zeldovich control variates (ZCV \cite{Kokron_2022,DeRose_2023_ZCV,Hadzhiyska_2023,shiferaw2024uncertaintiesgalaxyformationphysics,zhou2025csstcosmologicalemulatoriii}).

\begin{figure}
    \centering
    \includegraphics[width=\linewidth]{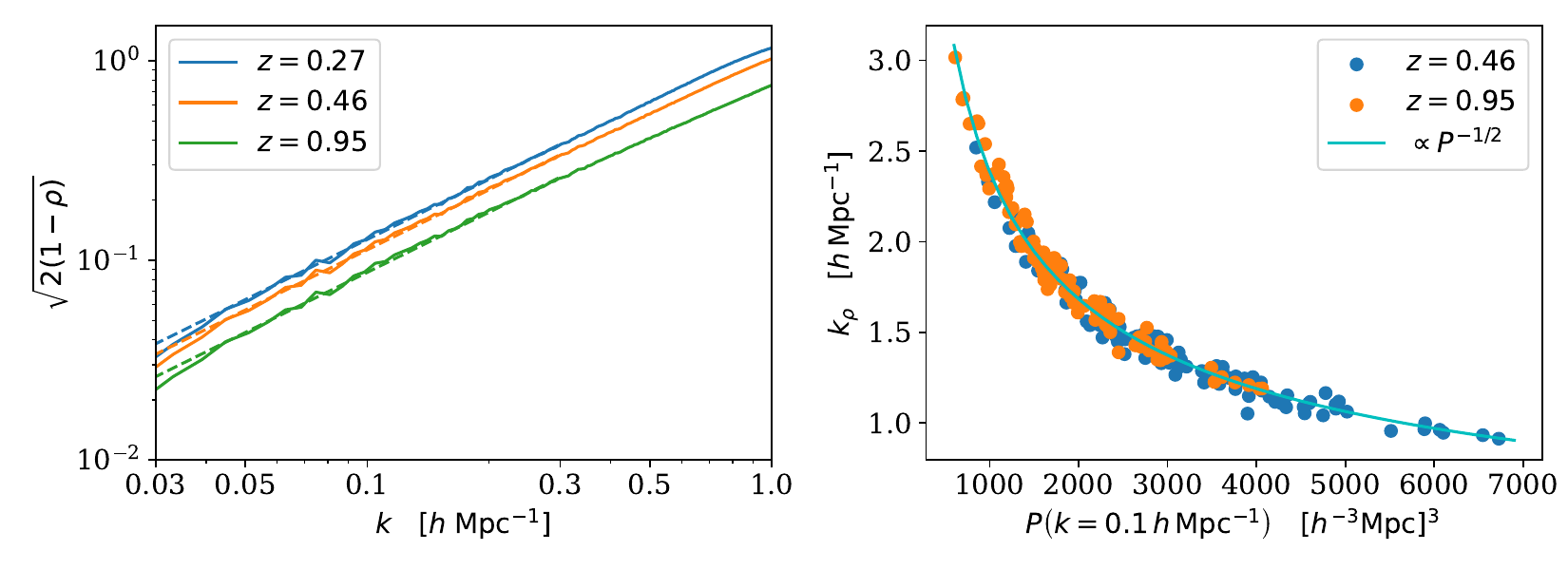}
    \caption{{\it Left}: $\sqrt{2(1-\rho)}$ for the Zeldovich and N-body matter fields as a function of $k$ from the first simulation in the Aemulus-$\nu$ suite for $z\approx 0.25$, 0.5, and 1.  This approximates, at low $k$, the reduction in standard deviation for $P_{mm}(k)$ due to employing ZCV.  The dashed lines show the fit to the form in the text, $1-\rho\approx (k/k_\rho)^2$. {\it Right}: The best-fit $k_{\rho}$ versus the measured $P_{\rm L}(k=0.1\,h\,{\rm Mpc}^{-1})$ for each Aemulus-$\nu$ cosmology at $z=0.46$ and $z=0.95$. The cyan curve represents the best-fit power law, with $k_{\rho} \propto P^{-1/2}$ jointly fitting the $z=0.46$ and $z=0.95$ points.}
    \label{fig:krho_fit}
\end{figure}

In our case, we combine our N-body estimate of the power spectrum, $P^N$, with the Zeldovich power spectrum, $P^Z$, to form $P^{ZCV}=P^N-\beta(P^{Z}-\langle P^{Z}\rangle)$.  At low $k$, the N-body and Zeldovich fields are highly correlated and ${\rm Var}[P^N(k)] \approx {\rm Var}[P^{Z}(k)]$.  For such scales, the uncertainty on the N-body $P(k)$ is reduced by $\sqrt{1+\beta^2-2\beta\rho}$ where $\rho$ is the cross-correlation coefficient between $P^N$ and $P^Z$.  For the optimal choice of $\beta=\rho$, where  the error on $P$ is reduced by $\sqrt{1-\rho^2}$ but motivated by ref.~\cite{Kokron_2022} we use a phenomenological form: $\beta=(1/2)\{1-\tanh[(k-k_0)/\Delta_k]\}$.  For low $k$, when $\beta$ and $\rho$ are both very close to 1, the error is reduced by $\approx\sqrt{2(1-\rho)}$ and approximating $1-\rho\approx (k/k_{\rho})^2$, with ``decorrelation scale'' $k_\rho$ (see figure~\ref{fig:krho_fit}), we have $ \Delta P_m/P_m \propto 1/\sqrt{k_{\rho}^2\,\Delta k}\,$ \cite{Kokron_2022}.
Compared to Eq.~(\ref{eqn:dPonP}), the variance is reduced by $k/k_{\rho}$ and now grows only as $k^{-1/2}$ for log-spaced bins in $k$ (it is constant for linearly spaced $k$ bins).

\begin{figure}
    \centering
    \includegraphics[width=0.8\linewidth]{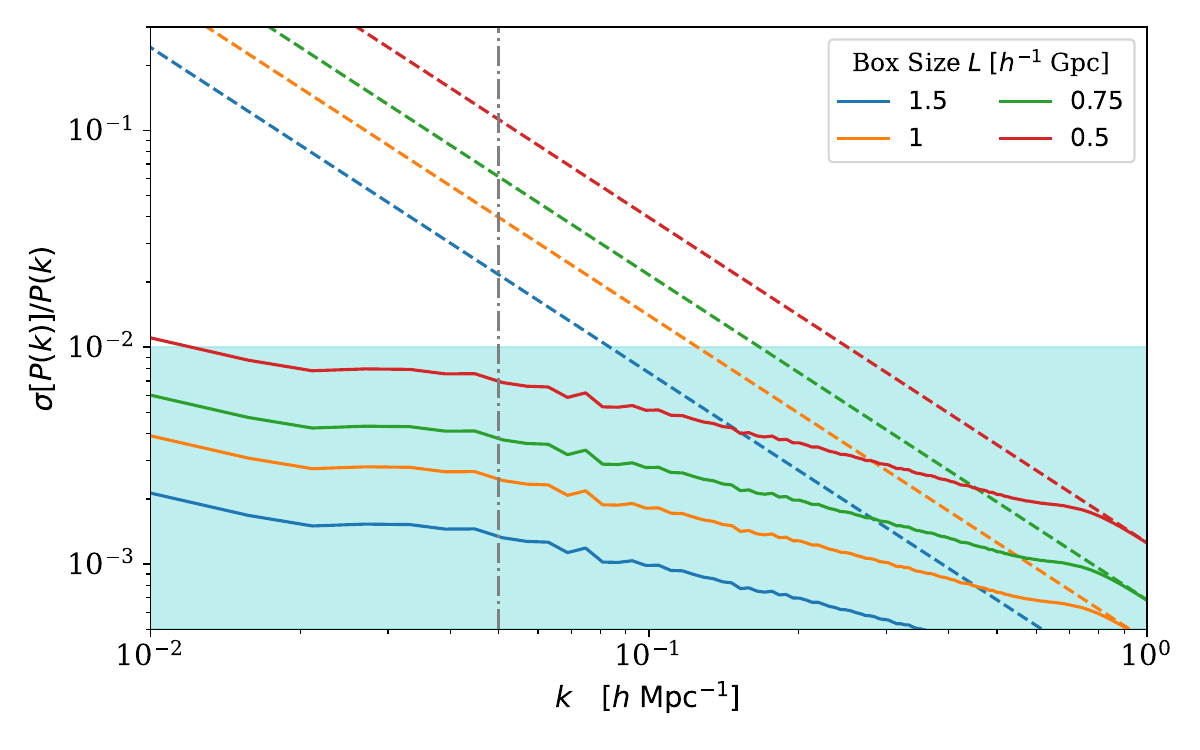}
    \caption{Fractional matter power spectrum error, at $z=0.5$, for various box sizes due to sample variance without ZCV (dashed) and with ZCV (solid) assuming $\Delta k/k=0.2$.  The vertical dot-dashed line shows $k=0.05\,h\,\mathrm{Mpc}^{-1}$, where we can reliably switch to PT.}
    \label{fig:Pmm_ZCV}
\end{figure}

Figure \ref{fig:Pmm_ZCV} shows the fractional error on the (matter) power spectrum for different choices of box size, with and without the introduction of ZCV.  The dashed lines show the ``raw'' error on $P$, following Eq.~(\ref{eqn:dPonP}).  One sees that the error becomes significant at $k\simeq 0.1\,h\,\mathrm{Mpc}^{-1}$, which can introduce noise in any simulation-calibrated emulator or fitting function unless either multiple boxes, or very large volumes in a single box, are run.  The introduction of ZCV tames this error considerably, such that it is manageable for boxes of $1\,h^{-1}$Gpc or more to $k\ll 0.1\,h\,\mathrm{Mpc}^{-1}$, below which we can safely switch to perturbation theory.  A similar argument holds for all of the other spectra that we need to simulate.

Another factor to consider when determining a simulation box size is the impact that missing modes larger than the box have on modes inside the box. The absence of these modes in the simulation leads to missing power for modes inside the simulation volume. Ref.~\cite{Klypin_2019} estimated the variance of the amplitude of power missed due to a finite, periodic simulation box as
\begin{equation}
    \sigma^2_{\rm miss}(L) \simeq \int_0^{k_{\rm box}}\frac{k^2\,dk}{2\pi^2}\ P_{\rm L}(k)
    \, , \quad
    k_{\rm box}= \frac{2\pi}{L} \, .
\end{equation}
For $z\ge 0$, this missing power is $\sigma_{\rm miss}<10^{-2}$ for a $L\ge 1\,h^{-1}\,{\rm Gpc}$ simulation box for cosmologies close to the currently preferred models.

Ensuring that analytic and grid-based LPT for a given simulation match also places requirements on the scales that should be resolved, and thus the minimum box size (and mass resolution). Ref.~\cite{Kokron_2022} shows that, in practice, the smallest and largest configuration-space scales probed by a simulation with $N^3$ particles and side length $L_{\rm box}$ are $q_{\rm min} \approx L_{\rm box}/N$ and $q_{\rm max}\approx L_{\rm box}/\pi$. The integrands for the first three Lagrangian basis spectra, $P_{11},\,P_{\delta 1}$, and $P_{\delta \delta}$, which make the largest contributions to $P(k)$ have significant support on scales as small as $q_{\rm min} \approx 1h^{-1}\,{\rm Mpc}$ and as large as $q_{\rm max} \approx 160\,h^{-1}\,{\rm Mpc}$ as shown in ref.~\cite{Kokron_2022}. For this pair of $q_{\rm min}$ and $q_{\rm max}$, simulations should be run in a box with minimum size $L_{\rm box}=500\,h^{-1}\,{\rm Mpc}$ with at least $N^3=(500)^3$ particles.  This is consistent with the ``missing modes'' arguments above.

\subsection{Mass resolution and particle load}

When determining the mass resolution/particle loads to use in a simulation, there are a number of trade-offs between higher- and lower-resolution simulations to consider. On the one hand, particle discreteness effects decrease as the number of particles increases. Additionally, the greater the mass resolution, the smaller the structures that can be resolved.  The major drawback of running higher-resolution simulations is the increased computational cost.  Most modern N-body simulation codes have $O(N\;{\rm log}N)$ complexity \cite{vogelsberger2019cosmologicalsimulationsgalaxyformation} and storage requirements, the latter of which usually is the biggest limitation on running very large simulations.  Large simulations require vast amounts of (expensive) disk space, access to computers with sufficient memory, and produce logistical challenges involved with using very large simulation outputs.

For standard simulation-based models of galaxy clustering statistics, such as HODs, the mass resolution requirements are typically set by the need to resolve the least massive halos hosting the galaxy samples of interest. For example, in DESI, emission line galaxies are hosted by halos with $M_{\rm halo}\sim10^{11}\, h^{-1}M_{\odot}$ \cite{Rocher2023}, and if one requires of order 200 particles to resolve a halo \cite{Warren2005,DeRose2019}, then that places a very onerous requirement that the particle mass must be less than $5\times10^{8}h^{-1}M_{\odot}$. Running simulations with these mass resolutions in box sizes $L=1\, h^{-1} \rm Gpc$ as discussed in the previous section requires approximately $5000^3$ particles, making such simulations extremely expensive. For reference, the AbacusSummit high-resolution simulation, run in a $L=1\, h^{-1} \rm Gpc$ box with $6300^3$ particles, required 1900 GPU-node hours on the Summit supercomputer at the Oak Ridge Leadership Computing Facility \cite{Maksimova_2021}. 

These requirements can be significantly relaxed by assuming a hybrid effective field theory-based model. Such models do not place any direct requirements on halo statistics, instead only requiring that the spectra entering into the HEFT bias expansion be resolved to the $k_{\rm max}$ of interest, and that the simulation be large enough so that these spectra agree with their LPT counterparts at the wavenumber where one wishes to transition to analytic calculations. In practice, these requirements can be met with a particle mass of order $10^{10}\, h^{-1}M_{\odot}$ \cite{DeRose_2023,DeRose2019} to resolve the matter power spectrum to $k\sim 1\, h\rm Mpc^{-1}$ at sub-percent accuracy, and a box size of $L=1\, h^{-1} \rm Gpc$ in order to transition to LPT at $k=0.05\, h\rm Mpc^{-1}$ \cite{Kokron_2022}. With standard time-stepping methods, and computing small-scale forces using an oct-tree algorithm as implemented in \texttt{Gadget-3}, a simulation with $1400^3$ particles in a $(1\, h^{-1}\rm Gpc)^3$ volume requires $\mathcal{O}(10^2)$ CPU-node hours on the Perlmutter machine at NERSC.

Even this may overestimate the required resources.  Much of the expense above is driven by time-stepping requirements to maintain accuracy on scales well below the mesh scale, required to resolve dark matter halos. If one is willing to forgo accuracy on sub-mesh scales, more sophisticated time-stepping algorithms can drastically increase the computational efficiency of simulations. For example, state-of-the-art time stepping schemes for particle mesh simulations can achieve $\ll 1\%$ converged results on the matter power spectrum up to $k\simeq 0.6 h\, \rm {Mpc}^{-1}$ in ten time steps using $384^3$ particles and a mesh of size $768^3$ in a simulation with side length $2\, h^{-1}\rm{Gpc}$ \cite{Rampf_2025,DISCO-DJ}.  While we wish to work to higher $k$ than $0.6 h\, \rm {Mpc}^{-1}$, $2h^{-1}{\rm Gpc}$ boxes are likely unnecessary. Reducing the box size to $1\,h^{-1}{\rm Gpc}$ and running a similar number of particles and mesh points, possibly with slightly more time steps, would satisfy our needs.  Such simulations can be run in a matter of minutes on a single GPU node at most modern supercomputing facilities, and given the requirements set down in this work, may provide the optimal solution for running simulations for HEFT models.

\subsection{Initial conditions}

An important factor to consider when initializing N-body simulations is the starting redshift. Discretization error and numerical artifacts arising from decaying modes suggest that simulations should be initialized as late as possible \cite{Crocce_2006, Angulo_2022}. However, perturbation theory is frequently used to generate initial conditions for simulations. If simulations are started too late, the perturbation theory may not apply over a sufficient range of scales. Ref. \cite{Michaux_2020} showcases that initializing at the latest possible times with higher-order perturbation theory yields the lowest error --- they find the lowest error in all summary statistics when initializing with 3rd-order LPT at $z=11.5$. This technique was utilized in ref.~\cite{DeRose_2023}.



The accuracy of particular starting times and LPT orders also depends on the resolution of the simulation: higher-resolution simulations require earlier starting times at fixed-order LPT because they resolve smaller scales that become nonlinear earlier. Convergence tests were performed on simulations with varying starting redshifts and resolutions in ref.~\cite{DeRose_2023}. They found that for simulations of volume $(525\;h^{-1}\;{\rm Mpc})^3$ evolving $700^3$ $cb$ particles and $700^3$ neutrinos on a particle mesh of size $1050^3$ initializing at $z=12$ with third-order LPT yields $<1\%$ error in the real-space, matter power spectrum for $k<4\;h\;{\rm Mpc}^{-1}$ at $z=0$, comfortably below our requirements.

\subsection{Summary}
\label{sec:nbody_summary}

The use of simulations for theoretical predictions requires that we run appropriate simulations at a number of different cosmologies.  In the next section, we discuss how to choose the grid of cosmologies at which simulations are run to ensure accurate emulation.  In this section, we described the requirements on the simulations to be run at each cosmology.  To summarize:

\begin{itemize}
    \item Simulations need to be run in boxes that are sufficiently large such that sample variance is small, missing power from modes larger than the simulation box is small, and analytic and grid-based LPT agree on the scales of interest. For a cubic box with side length $L_{\rm box}=1\,h^{-1}{\rm Gpc}$, $\sigma[P(k)]/P(k) < 0.3\%$ for $k$ between $0.01$ and $1\,h\,{\rm Mpc}^{-1}.$ The missing power due to missing super-box modes is also $<1\%$.  The simulation volume and resolution should be such that the analytic and grid-based LPT agree on the scales of interest. The first three Lagrangian basis spectra integrands have support in the range $q \in \sim[1, 160]\,h^{-1}\,{\rm Mpc}$ for $k \in [0.05,1]\,h\,{\rm Mpc}^{-1}$, compatible with $L\simeq 1\,h^{-1}$Gpc and $1000^3$ particles.
    \item Initializing simulations with higher-order LPT at lower redshift yields more-converged results compared to simulations initialized earlier at lower-order LPT \cite{DeRose_2023, Michaux_2020}. In line with those findings, we recommend initializing simulations at $z=12$ with third-order LPT when using the resolutions recommended in this section.
    \item With standard time stepping criteria, and computing small-scale forces using an oct-tree as implemented in \texttt{Gadget-3}, a simulation with $1400^3$ particles in a $(1\, h^{-1}\rm Gpc)^3$ volume requires $\mathcal{O}(10^2)$ CPU-node hours on the Perlmutter machine at NERSC.  Resolving halos, however, is unnecessary for HEFT.  If one is willing to sacrifice accuracy on sub-mesh scales, more sophisticated time-stepping algorithms and lower force resolution can greatly accelerate simulations.  Under these conditions, $\le 1\%$ converged results on the scales of interest can be achieved in a matter of minutes on a single GPU node.  This allows a very large number of simulations to be run, reducing errors in interpolation.
\end{itemize}

\section{Emulator}
\label{sec:Emulator}

Running $N$-body simulations for every cosmology in, say, an MCMC analysis would be prohibitively expensive. However, the response of the nonlinear power spectrum to changes in cosmological parameters is smooth. This means that one can interpolate between cosmologies using an emulator. In previous sections, we examined how to obtain accurate predictions at specific points in our parameter space. We now aim to understand the emulator-induced interpolation error. To examine this without needing to run a very large number of N-body simulations, we use a combination of perturbation theory and \texttt{HMcode2020} \cite{Mead_2021} as a stand-in for the simulations.  Though there are some subtleties that we will discuss later, this combination varies with parameters in a manner very similar to the true matter power spectrum \cite{DeRose_2023, EuclidEmulator_2021} and thus is an appropriate stand-in for our purposes. 

We note that our emulator requirements will be set just by the matter power spectrum $P_{11}(k)$. We do so because, on large scales, the dominant basis spectra scale with changes in cosmology like $P_{\rm lin}(k)$ and hence $P_{11}$. On nonlinear scales, most of the basis spectra have similar sensitivity to changes in cosmology, and those that do change more rapidly with cosmology are relatively small contributions to the overall power spectrum, as shown in figures 10 and 11 of ref.~\cite{DeRose_2023} and which we have independently verified.

We created an emulator to predict the ratio of the nonlinear matter power spectrum to the corresponding one-loop LPT predictions for a given set of cosmological parameters. We use a combination of principal component analysis (PCA), for dimensionality reduction and polynomial chaos expansion (PCE), which approximates the output of a complex system as a sum of orthogonal polynomials with random coefficients. Further details are discussed in Appendix \ref{app:emu_constr}. We do not emulate the matter power spectrum itself, but the logarithm of the ratio of the model predictions to the 1-loop matter power spectrum
\begin{equation}
    \Gamma(k,\Omega)={\rm log}_{10}\left(\frac{P_{\rm model}(k,\Omega)}{P_{\rm 1-loop}(k,\Omega)}\right),
\label{eqn:Gamma_defn}
\end{equation}
where $\Omega$ is the set of $w_0w_a$CDM+$m_\nu$ cosmological parameters plus $\sigma_8(z),$ which we use as our time parameter in place of redshift. Both dividing by $P_{\rm 1-loop}$ and taking the logarithm of the ratio serve to reduce the dynamic range we must emulate. Our reasoning for emulating ${\rm log}(P_{NL}/P_{\rm 1-loop})$, as opposed to the more commonly emulated (log of the) nonlinear correction factor (NLC) $P_{NL}/P_{\rm lin}$, is because the former has a smaller dynamic range and more accurately accounts for the broadening of the baryon acoustic oscillations, as discussed in Appendix \ref{app:ratio}. To train the emulator, we first generate linear matter power spectra for 200 cosmologies at 30 logarithmically-spaced redshifts $z \in [0,3]$ for 700 $k$s using \texttt{CLASS} \cite{Diego_Blas_2011}. The nonlinear and one-loop LPT power spectrum predictions are then produced using \texttt{HMcode2020} and \texttt{velocileptors}, respectively. The degree to which HMcode is an appropriate proxy for simulation measurements over the scales in question is discussed in depth in Appendix \ref{app:ratio}.

\begin{figure}[htbp]
    \centering
    \includegraphics[width=\linewidth]{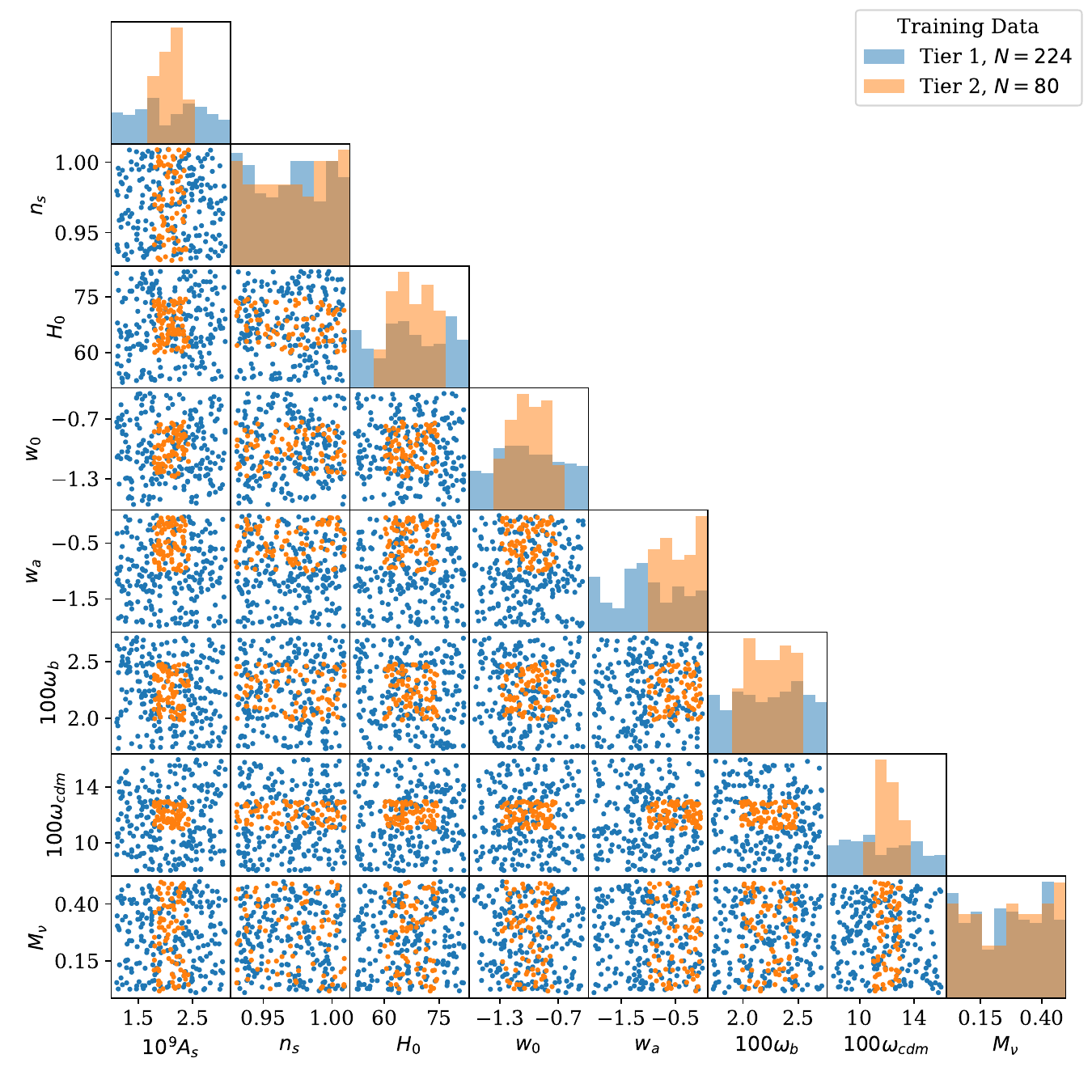}
    \caption{Locations in the 8D parameter space of the simulations for tier 1 (blue; $N=224$) and tier 2 (orange; $N=80$) samples.}
    \label{fig:corner_plot_tiers}
\end{figure}

Several of these training sets were generated and are compared in section \ref{sec:Results}. Gaussian noise is artificially added to the \texttt{HMcode2020} spectra, to mimic the sample variance from a finite simulation volume, appropriately reduced via ZCV. The cosmologies are selected quasi-randomly using Latin hypercube sampling, maximizing the minimum distance between pairs of points in two dimensions. The 8-dimensional parameter space being sampled consists of $\{A_s, n_s, h, w_0, w_a, \omega_b, \omega_{cdm}, \sum m_{\nu}\}$. The volume of this parameter space varies between training sets. We use two ``tiers'' of simulations focused on covering a broad parameter space at lower fidelity and a narrower space close to the models preferred by current data at higher fidelity. The minimum and maximum values of each parameter for the different ``tiers'' are given in table \ref{tab:tier_bnds_table}. The locations of the minimum training set samples in table~\ref{tab:N_train} are shown in figure~\ref{fig:corner_plot_tiers}.

    \begin{center}
    \begin{table}
        \centering
        \begin{tabular}{|c||c|c|c|c|}
            \hline\hline
            & Tier 1 min. & Tier 1 max. & Tier 2 min. & Tier 2 max. \\
           \hline
           $10^9A_s$ & 1.10 & 3.10 & 1.77 & 2.43 \\
           $n_s$ & 0.93 & 1.01 & 0.93 & 1.01 \\
           $H_0$ & 52.0 & 82.0 & 59.5 & 74.5 \\
           $w_0$ & -1.56 & -0.44 & -1.28 & -0.72 \\
           $w_a$ & -2.0 & 0.0 & -1.0 & 0.0 \\
           $\omega_b$ & 0.0173 & 0.0272 & 0.0198 & 0.0248 \\
           $\omega_c$ & 0.08 & 0.16 & 0.11 & 0.13 \\
           $\sum m_{\nu}$ & 0.01 & 0.50 & 0.01 & 0.50 \\
           \hline
        \end{tabular}
        \caption{Tier 1 and 2 parameter space bounds for emulator training sets. Tiers 1 and 2 are identical to those used in \cite{DeRose_2023}, except for the bounds on $w_a$. Further details regarding how these bounds were determined are given in section \ref{sec:Emulator}.
        }
        \label{tab:tier_bnds_table}
    \end{table}
    \end{center}
    
Tier 1 and 2 training sets were also generated for $\Lambda$CDM, $\Lambda$CDM + massive neutrinos ($m_{\nu}$), and $w$CDM+$m_{\nu}$ with parameter space dimensionalities of 5, 6, and 7, respectively. These were generated with the aim of finding a relationship between the parameter space dimensionality $D$ and the size of the training dataset $N$ required for emulator predictions to achieve some accuracy. We additionally expect the degree to which the matter power spectrum changes with each cosmological parameter to affect the parameter space bounds. 

\section{Results and worked examples}
\label{sec:Results}

Following ref.~\cite{DeRose_2023}, tier 1 spans as wide a parameter space as possible. The required number of simulations listed in table \ref{tab:N_train} for tier 1 is approximately the minimum number required to achieve $<2\%$ error in the 68th percentile for $k<1\;h^{-1}$Mpc. Tier 2 is a smaller parameter space, where we expect regions of high likelihood to lie, with the number of simulations in the table being the number required such that the emulator error is $<1\%$ for the 68th percentile for $k<1\;h^{-1}$Mpc.

\begin{center}
\begin{table}[]
    \centering
    \begin{tabular}{|c||c|c|}
            \hline\hline
            Cosmology & Tier 1 N & Tier 2 N \\
           \hline
           $w_0w_a{\rm CDM}+m_{\nu}$ & 224 & 80 \\
           $w{\rm CDM}+m_{\nu}$ & 150 & 80 \\
           $\Lambda{\rm CDM}+m_{\nu}$ & 96 & 36 \\
           $\Lambda$CDM & 60 & 24 \\
           \hline
    \end{tabular}
    \caption{Table showing the minimum required training set size given our error goals for tiers 1 and 2. We present the values for $w_0w_a$CDM+$m_\nu$, $w$CDM+$m_\nu$, $\Lambda$CDM+$m_\nu$, and $w_0w_a$CDM to showcase the scaling between the required number of training simulations and the number of free model parameters.} 
    \label{tab:N_train}
\end{table}
\end{center}

As mentioned in the previous subsection, these analyses were also done for $\Lambda$CDM, $\Lambda$CDM + massive neutrinos, and $w$CDM + massive neutrinos with tier 1 and tier 2 parameter space bounds. The minimum number of training spectra required for our desired error budgets for tiers 1 and 2 is given in table~\ref{tab:N_train}.

An alternative view is shown in table \ref{tab:percentile_error}.  Here we list the $68^{\rm th}$ and $95^{\rm th}$ percentile error on the matter power spectrum at $k=0.3\,h\,\mathrm{Mpc}^{-1}$ and $k=0.6\,h\,\mathrm{Mpc}^{-1}$ as a function of the number of training points for both tier 1 and tier 2.  This gives an indication of how the error degrades as we increase the volume of the parameter space.  We focus on the full 8-dimensional parameter space, where the differences are most pronounced.  Table \ref{tab:percentile_error} additionally shows how the tails of the error distribution behave with the number of training samples and the dimension of the space.

\begin{center}
\begin{table}[]
    \centering
    \begin{tabular}{|c||cccc|cccc|}
    \hline\hline
    & \multicolumn{4}{c|}{$68^{\rm th}$ percentile} & \multicolumn{4}{c|}{$95^{\rm th}$ percentile} \\
    & \multicolumn{2}{c}{Tier 1} &  \multicolumn{2}{c|}{Tier 2} & \multicolumn{2}{c}{Tier 1} &  \multicolumn{2}{c|}{Tier 2} \\
    Number & 0.3 & 0.6 & 0.3 & 0.6 & 0.3 & 0.6 & 0.3 & 0.6 \\
    \hline
    100    & 0.8\% & 2.5\% & 0.5\% & 0.9\% & 2.3\% & 6.8\% & 1.2\% & 1.9\% \\
    200    & 0.7\% & 2.1\% & 0.2\% & 0.3\% & 2.2\% & 6.7\% & 0.5\% & 1.3\% \\
    400    & 0.3\% & 1\% & 0.08\% & 0.13\% & 1\% & 2.8\% & 0.2\% & 0.3\% \\
    \hline
    \end{tabular}
    \caption{The $68^{\rm th}$ and $95^{\rm th}$ percentile error at $k=0.3\,h\,\mathrm{Mpc}^{-1}$ and $k=0.6\,h\,\mathrm{Mpc}^{-1}$ as a function of the number of training points for both tier 1 and tier 2.  We quote results for the $w_0w_a$CDM+$m_\nu$ case to showcase the scaling between the required number of training simulations and the volume of the space.} 
    \label{tab:percentile_error}
\end{table}
\end{center}

To give further intuition, below we estimate the number of simulations required to train a HEFT emulator on our tier 1 and tier 2 parameter spaces for four use cases representative of analyses that might be done with current or future experiments.  Our cases are chosen to illustrate different scales and dominant statistics. For each, we estimate the relevant statistical errors and use this to infer the required number of training simulations such that the 68th percentile interpolator error is $\lesssim 0.5$ (tier 1) and $\lesssim 0.33$ (tier 2) the statistical error on all relevant scales. We employ both the Born and Limber approximations to convert between $P(k)$ and $C_\ell$. We have verified that the redshift interpolation error is sub-percent over the scales in question, and do not expect it to affect our forecasts.

\subsection{DES Y6 GGL with DESI BGS}
\label{sec:example_DES}

DES Y6 is currently the widest area, deep lensing sample that is publicly available. Additionally, the DESI bright galaxy survey (BGS) sample is also in the optimal redshift range to produce a GGL signal with DES Y6 sources. 
DESI BGS is a high-density sample, with $\bar{n}_{\theta}\approx 800\,{\rm deg}^{-2}$. Thus, the shot noise in this sample remains subdominant to $\ell\approx 900$, and is not a major contributor to the error on the scales of interest.  Additionally, DESI has obtained accurate spectroscopy for the BGS sample. As a result, 3D clustering may be used rather than projected 2D clustering. As shown in ref.~\cite{Maus25}, once 3D clustering is obtained, there is little to no gain from $C_\ell^{gg}$. For this reason, we focus on forecasts for $C_{\ell}^{\kappa g}$. 

\begin{figure}
    \centering
    \includegraphics[width=\linewidth]{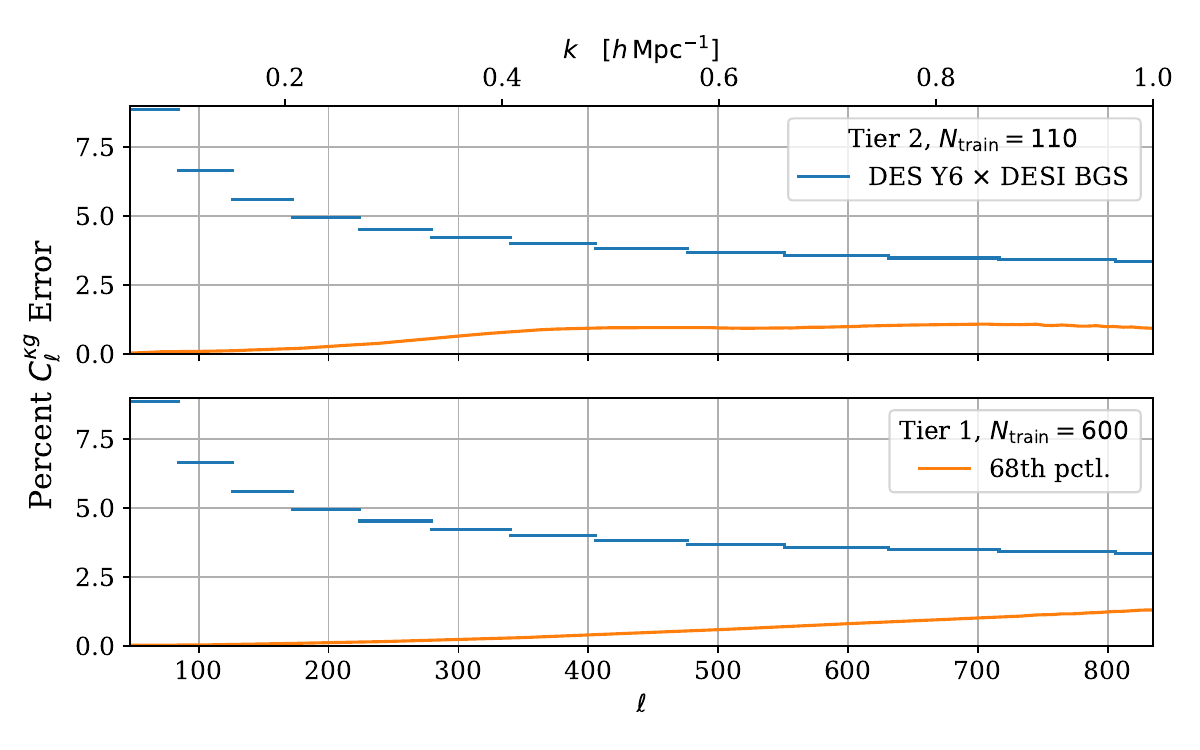}
    \caption{Emulator error for our tier 2 (top) and tier 1 (bottom) parameter space compared to statistical $C_{\ell}^{\kappa g}$ error. For our lens sample we use DESI BGS, with $\mu_z=0.3$, $\sigma_z=0.2$, $b=1$, and $\bar{n}_{\theta}=800\,{\rm deg}^{-2}$. For the source sample, we use the highest-$z$ DES Y6 source bin, and take $n_{\rm eff}=1.6\,{\rm arcmin}^{-2}$, $\sigma_{\epsilon}=0.26$, and $f_{\rm sky}=0.1$. We include a secondary $k-$axis, with $k=(\ell+0.5)/\chi(z=0.3)$ for orientation.
    }
    \label{fig:DES_GGL_res}
\end{figure}

The shape noise associated with the Y6 DES source sample exceeds the cosmological signal at $\ell \approx 180$, which limits the range of scales where precise measurements can be made. The resulting error in $C_\ell^{\kappa g}$ is highest ($\sim 9\%$) at  low $\ell$ and plateaus around $\sim 3\%$ at high $\ell$, as shown in figure~\ref{fig:DES_GGL_res}. Though DES Y6 is the widest deep survey currently available, the limited sky coverage $f_{\rm sky} \approx 0.1$ drives up the variance on all scales.

To train a $w_0w_a$CDM$+m_\nu$ emulator for DES Y6 $\times$ DESI BGS GGL analysis, we estimate that $\sim 600 $ and $\sim 110$ simulations will be required for a tier 1 and tier 2 emulator, respectively.  The 68th percentile emulator error that would result from such a simulation campaign is shown in figure~\ref {fig:DES_GGL_res}.  For the tier 1 sample, one sees a monotonic increase in the error to smaller scales (larger $\ell$).  For the tier 2 sample, with just 110 simulations in the training set, the error reaches the percent level at $\ell\approx 500$.

Similar conclusions apply to the DECADE weak lensing sample \cite{anbajagane2025DECADE_shear_I, Anbajagane2025}.  DECADE represents the first public wide-area shear survey.  With an area $3\times$ larger than DES, it is a natural precursor to the Euclid and LSST datasets we expect in future years.  The larger $f_{\rm sky}$ lowers the errors at low $\ell$.  However, DECADE has a relatively low source density, and thus becomes shape-noise limited even before DES, limiting the constraining power at high $\ell$.  The natural lens sample for GGL remains the DESI BGS or the lower redshift LRGs, since we ideally want the distance to the lenses to be $\approx 1/2$ the distance to the sources.  For this reason, 3D clustering is preferred over $C_\ell^{gg}$ and the galaxy density is high enough to be a negligible contribution to $\sigma[C_\ell^{\kappa g}]$ on all relevant scales.  Emulators for GGL with DECADE can be trained with the same number of simulations as for DES.

\subsection{SO CMB lensing with DESI LRGs}
\label{sec:example_so}

\begin{figure}
    \centering
    \includegraphics[width=\linewidth]{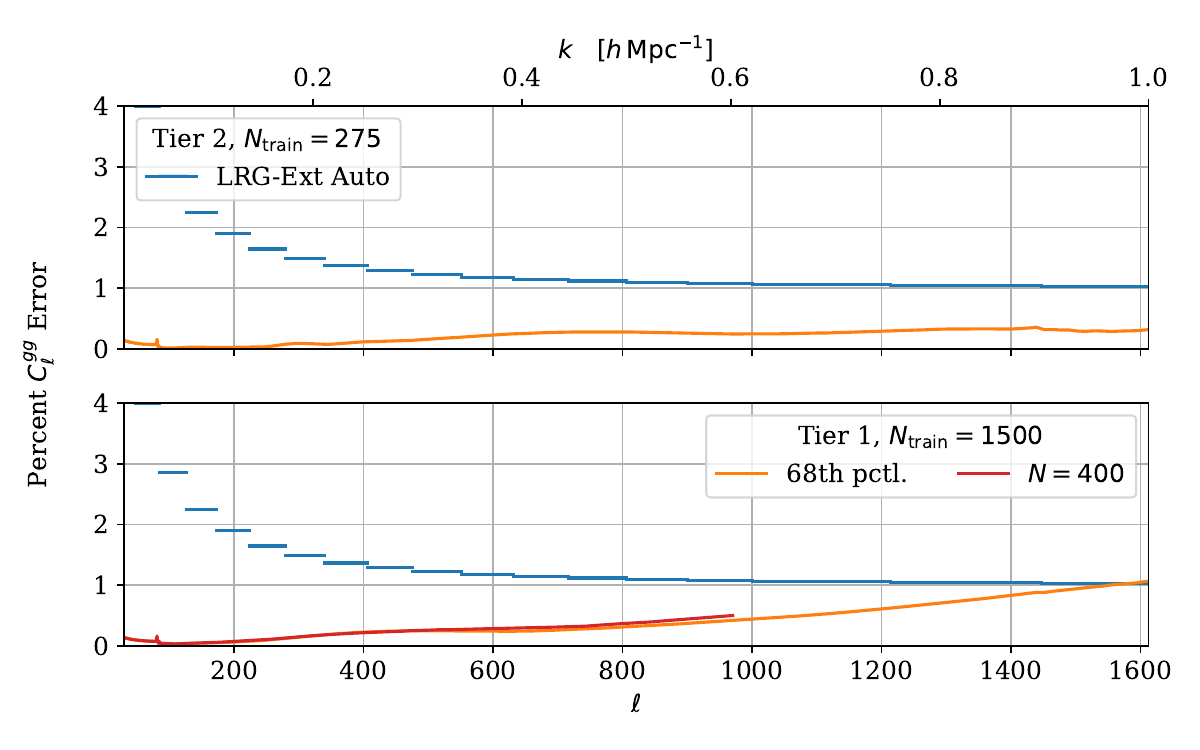}
    \caption{Emulator error for our tier 2 (top) and tier 1 (bottom) parameter space compared to statistical $C_{\ell}^{gg}$ error. We use our fiducial extended LRG-like sample, and take the DESI SO sky overlap to be $f_{\rm sky}=0.35$. The red curve in the bottom panel shows the error for an emulator trained to $k_{\rm max}=0.6\,h\,{\rm Mpc}^{-1}$, for which only 400 training points are required to meet our error goals. We include a secondary $k-$axis using $k=(\ell+0.5)/\chi(z=0.633)$ for orientation.
    }
    \label{fig:CMB_lens_res}
\end{figure}

The contributions to the statistical error in the CMB lensing galaxy cross-spectrum are the CMB lensing noise curve and the galaxy shot noise. As in section \ref{subsec:stat_errs}, we use the publicly available SO lensing noise curve, and galaxy redshift distribution and number densities consistent with an extended LRG-like sample. The Poisson shot noise for our extended LRG sample is subdominant to $\ell \approx 900$, while the SO lensing noise is only subdominant for $\ell \lesssim 380$. Since we do not have spectroscopy of an extended DESI LRG sample and thus cannot use 3D clustering, our emulator error goal is set by the statistical error on $C_{\ell}^{gg} $, which is small and rapidly approaches $1\%$ with increasing $\ell$. 
As shown in figure~\ref{fig:CMB_lens_res}, a large number of training simulations are thus required to meet our error goals. To train a $w_0w_a$CDM$+m_\nu$ emulator for SO $\times$ extended DESI LRGs, we estimate 275 simulations will be required for a tier 2 emulator. For tier 1, because of the very small statistical error in $C_{\ell}^{gg}$ at high-$k$, using even $1500+$ simulations only yields an emulator error $<50\%$ of the statistical error to $k\approx 0.7\,h\,{\rm Mpc}^{-1}.$ We note, however, that HEFT is likely not $0.5\%$ accurate (especially at high-$k$) and the higher $k$s contribute relatively little to our cosmological constraints due to the flexibility of the bias model.  If one wishes to emulate only to $k=0.6\,h\,{\rm Mpc}^{-1}$, we estimate that only 400 simulations are required.

\subsection{RST cosmic shear}
\label{sec:example_RST}

\begin{figure}
    \centering
    \includegraphics[width=\linewidth]{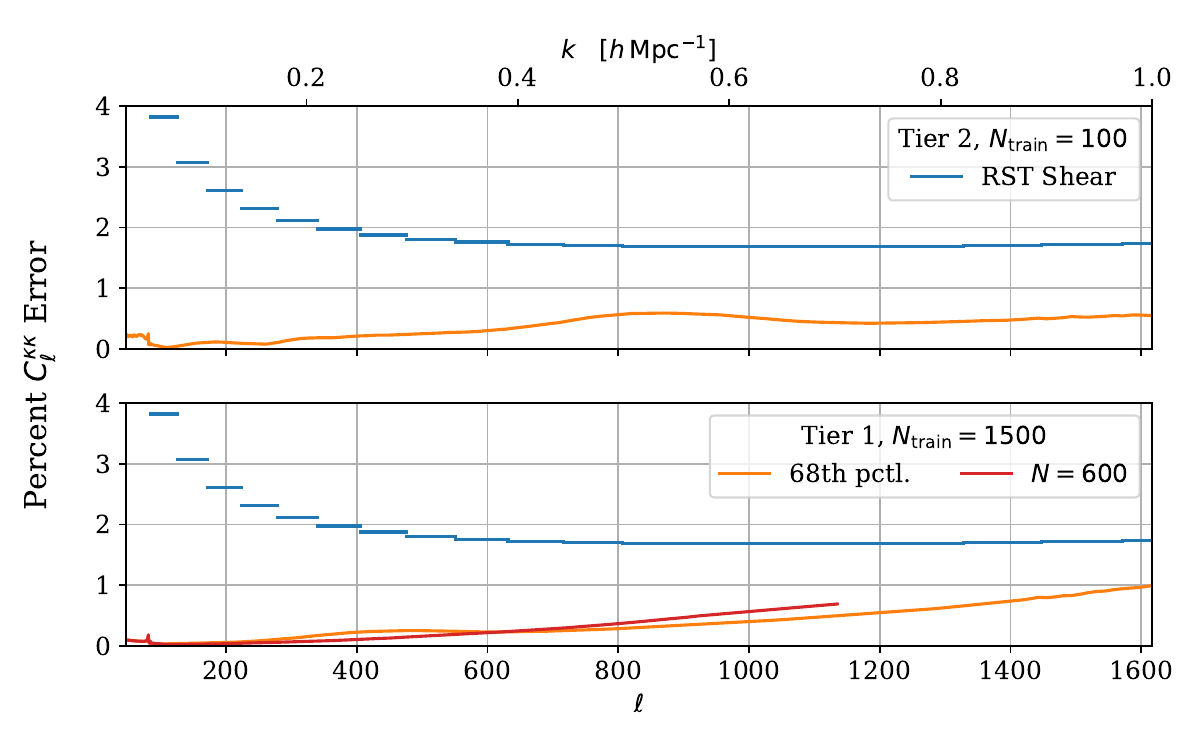}
    \caption{Emulator error for our tier 2 (top) and tier 1 (bottom) parameter space compared to statistical $C_\ell^{\kappa\kappa}$ error. We consider the combination of the RST HLWAS medium and wide tiers, as described in ref.~\cite {committee2025roman}. For our source $dN/dz$, we take the highest-$z$ source bin as described in figure~2 of ref.~\cite{Eifler_2021}. We include a secondary $k-$axis, with $k=(\ell+0.5)/\chi(z=0.635)$ for orientation.
    }
    \label{fig:RST_shear_res}
\end{figure}

The Roman Space Telescope high latitude wide area survey (RST-HLWAS) \cite{committee2025roman} provides an illustration of a very different sample, with very different requirements.  The combined medium and wide tiers of the RST-HLWAS achieve a high source density, and hence a low shape noise, but only over a very limited area of sky.  This means the statistical uncertainty at low $\ell$ is quite large, and the constraints are more heavily weighted to the higher $\ell$.  On these scales, the modeling is more complex, and it is also harder to achieve good emulator performance, making this an interesting case to consider.

The medium tier of the RST-HLWAS is expected to yield about 360 million galaxy shape measurements but only over a limited area, $f_{\rm sky}\approx 0.06$.  The wide tier is a disjoint measurement on an approximately equal-sized patch of sky, albeit with reduced number density.  We shall consider the combination of the two, with the resulting $C_\ell$ uncertainty obtained by inverse variance weighting that from each of the two surveys.
For this example, we consider the highest-$z$ source bin used in ref.~\cite{Eifler_2021} in the optimistic photo-$z$ case ($\sigma_z=0.01$). We take $n_{\rm eff}=4.1\,{\rm arcmin}^{-2}$ for this bin in the medium tier, $n_{\rm eff}=1.3\,{\rm arcmin}^{-2}$ in the wide tier, and fix $\sigma_{\epsilon}=0.26$.  As shown in figure~\ref{fig:RST_shear_res}, the uncertainty on $C_\ell^{\kappa\kappa}$ is large enough that the IA contribution to the autocorrelation of this particular slice is subdominant to the statistical error.  For this reason, and to illustrate a shear-only case, we concentrate on $C_\ell^{\kappa\kappa}$.  We do not include modeling uncertainties due to baryonic effects, and thus the requirements on the emulator are likely overestimated.

The combination of the medium and wide tiers yields a `low' statistical error at high $\ell$, as evident in figure~\ref{fig:RST_shear_res}.  These small scales are where the power spectrum changes more quickly with parameters and thus a large training sample is needed to achieve our error goals. To train a $w_0w_a$CDM$+m_\nu$ emulator for RST cosmic shear analysis, we estimate that $\sim 1500$ and $\sim 100$ simulations will be required for a tier 1 and tier 2 emulator, respectively.  As in the case of CMB lensing and DESI extended LRGs, the high number of simulations required to meet our tier 1 goals is driven primarily by the high-$k$ emulator error, which we regard as a ``soft requirement'' due to the other major sources of uncertainty that enter at high $k$.  The total ``cost'' is quite sensitive to the assumed $k_{\rm max}$.  If one wishes accurate emulation only up to $k_{\rm max}=0.7\,h\,{\rm Mpc}^{-1}$, assuming the modeling error will be dominated by other sources at higher $k$, only $\sim 600$ simulations are required (i.e.\ about half the cost).  These numbers do not represent a major commitment of resources if cheap (e.g.~PM) simulations are used. However, they would be substantial if high-resolution simulations were to be run at each point in parameter space.  This might be the case if one desired the simulations for other uses (e.g.\ other statistics or other science cases).  This suggests that the design of an emulator for surveys, such as RST-HWLAS, which depend sensitively on high-$k$ modeling, should be done in the context of the suite of other modeling choices that are likely to be made.

\subsection{A ``steel'' sample}
\label{sec:steel}

As discussed in some detail in section \ref{sec:Motiv_Acc_Req}, uncertainties in baryonic feedback and scale-dependent bias limit our ability to extract cosmological information from measurements at small scales.  At the same time, errors associated with intrinsic alignments or calibration of $dN/dz$ through photometric redshifts can quickly dominate over statistical uncertainties in the Stage IV era.  However, the first point suggests a means of addressing the second.

For a fixed number of sources, lenses, or galaxies, the shape of the power spectrum implies the SNR per mode is higher on quasi-linear scales than on very small scales (with $d\ln P(k)/d\ln k\simeq -2$ near $k\simeq 0.2\,h\,\mathrm{Mpc}^{-1}$).  If our cosmological information comes primarily from such larger scales, which are closer to sample variance limited, then we lose little information by working with lower source, lens, or galaxy number densities.  It makes sense then to ``cherry pick'' the subset of objects where the systematic errors are least, so as to reduce the combination of statistical and systematic errors that ultimately matters for our inferences \cite{SteelSample}.

In this section, we consider what our emulator performance would have to be in order to model such a ``steel sample'' of galaxies and lenses \cite{SteelSample}.  Specifically, we assume $dN/dz$ shapes as for the LSST gold sample, dropping the highest $z$-bin, but with a much reduced number density.  For illustration, we take 1 galaxy per arcmin${}^2$.  We imagine that this subset of objects has been chosen such that it is possible to spectroscopically follow a large number of them up so as to calibrate $dN/dz$ and provide priors on IA from direct measurements of the sample.  Under these conditions, the statistical power of the steel sample is comparable to that of the LSST gold sample once uncertainties in modeling are taken into account \cite{SteelSample}.

\begin{figure}
    \centering
    \includegraphics[width=\linewidth]{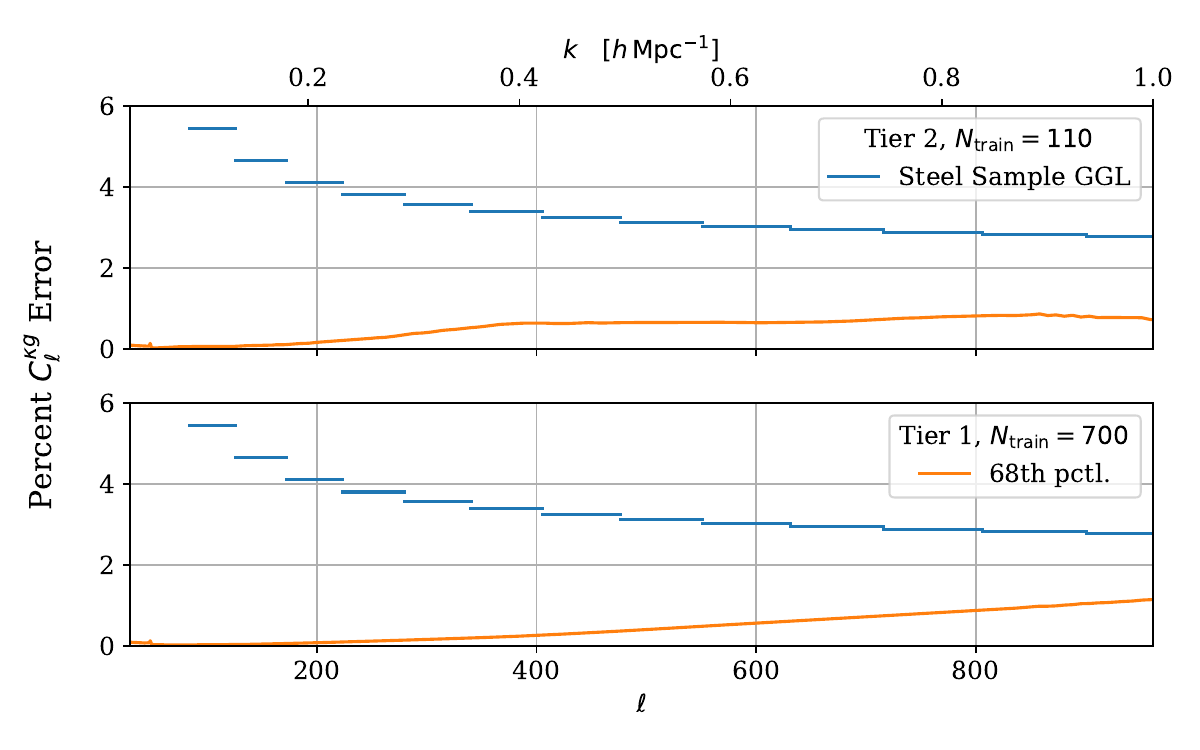}
    \caption{Emulator error for our tier 2 (top) and tier 1 (bottom) parameter space compared to statistical $C_\ell^{\kappa g}$ error for the steel sample, described in the text of section \ref{sec:steel}. We include a secondary $k-$axis, with $k=(\ell+0.5)/\chi(z=1.65)$ for orientation.
    }
    \label{fig:steel_GGL_res}
\end{figure}

As in the case of DES Y6 GGL with DESI BGS described in section \ref{sec:example_DES}, because we will have spectroscopy for these galaxies, our error goals are set by the statistical error in $C_{\ell}^{\kappa g}$. The relatively low number density of the source sample means the shape noise for this sample exceeds $C_{\ell}^{\kappa \kappa}$ for $\ell \gtrsim 150$. The resulting error in $C_\ell^{\kappa g}$, shown in figure~\ref{fig:steel_GGL_res} peaks at about $8\%$ at low $\ell$, and decreases towards $3\%$ with increasing $\ell$.
To achieve our error goals, 700 and 110 training simulations are required for tier 1 and tier 2, respectively.

\section{Conclusion}
\label{sec:conclusions}

Our most robust cosmological constraints from projected clustering statistics at present come from hybrid models that combine elements of analytic and numerical approaches.  This, in turn, requires training emulators on grids of numerical simulations. In this paper, we have attempted to forecast the cost of generating such training data, making informed assumptions about the degree of statistical and systematic errors we expect from upcoming surveys and the range of scales over which we require robust predictions.

By studying the statistical errors expected to be delivered by current and future surveys, and accounting for sources of modeling uncertainty such as photometric redshifts, intrinsic alignments, scale-dependent bias, the impact of baryonic physics and non-linear evolution, we set a goal for our theoretical models of 1-2\% precision on scales $k<1\,h\,\mathrm{Mpc}^{-1}$.  We argue that achieving this performance is entirely within reach, even for 8-dimensional parameter spaces that include evolving dark energy and massive neutrinos.

We first validate the hybrid approach for an extended set of cosmological models, including models with (rapidly) evolving dark energy.  We confirm earlier findings of the range of validity of the model by studying the inclusion of a cubic bias operator, demonstrating that it impacts only small scales (as expected).  On very small scales, our errors are dominated by the complex physics being modeled, and our hybrid approach breaks down.  Given our current uncertainties in baryonic feedback and the scale-dependence of galaxy bias, we show that a wide range of cosmological models can fit the same small-scale clustering data.  Our most robust cosmological constraints thus come from quasi-linear scales, partially negating any advantages from improved emulator performance at high $k$.  At large scales and high redshift, our demands can easily be met by analytic models based upon cosmological perturbation theory.  Thus, we focus on obtaining accurate, simulation-based predictions for the range $0.05<k<1\,h\,\mathrm{Mpc}^{-1}$ and $0<z<2$.

The above goals can be translated into requirements on simulation volume, resolution, starting redshift, and data extraction.  We find that simulations should employ $\sim 10^9$ particles in a box of side no less than $1\,h^{-1}$Gpc.  The initial conditions of simulations at this resolution should be generated at $z\approx 12$, with third-order LPT.  Significant savings in computational cost can be achieved by the use of large, global time steps and coarse force resolution with little impact upon the spectra of interest for hybrid modeling.  We leave detailed investigations of the optimal configuration and the possibilities of GPU acceleration to a future publication.

Lacking a large grid of N-body simulations to investigate the training of emulators, we use a surrogate model.  We replace the prediction of the matter power spectrum from an N-body simulation with a combination of HMcode and perturbation theory.  The predictions of this model vary with cosmological parameters in a way very similar to the spectra from  ``true'' N-body simulations, thus providing a decent model of the response surface of our emulated quantities.  We can then generate large amounts of ``training data'' quite cheaply, in order to investigate how many simulations would be required to achieve a given performance over a certain parameter space.  We present some topical examples for current and future surveys in cosmologies with massive neutrinos and dynamical dark energy (8 cosmological parameters).

For combinations of experiments where 3D clustering information is available (e.g., \ galaxy redshift surveys) and only the cross-spectrum between lensing and galaxies drives the emulator requirements, we find that as few as 100 simulations are sufficient.  The inclusion of the shear or convergence auto-spectrum does not significantly alter this finding at the levels of Stage III experiments or for the ``steel sample'', though a shear-only analysis of a full stage IV sample would require more. The China Space Station Telescope (CSST) Cosmological Emulator \cite{zhou2025csstcosmologicalemulatoriii} is a currently available $w_0w_a$CDM $+m_{\nu}$ emulator that is $\sim 1\%$ accurate for $k \leq 1\,h\,{\rm Mpc}^{-1}$ and $0 \leq z \leq 3$ for their chosen parameter space, that would be well-suited for these cases. The Aemulus-$\nu$ emulator is also well-suited for this task, but is trained only over $w{\rm CDM}+m_{\nu}$ cosmologies.

The most demanding case arises when our requirements are set by modeling $C_\ell^{gg}$ for high-density samples.  In these cases, the number of simulations required to train the emulator can rise to over 1000. This would be extremely expensive if high-resolution N-body simulations were required; however, it remains quite a modest expense if particle mesh simulations with global timesteps are used instead. Current $w_0w_a$CDM$+m_{\nu}$ HEFT emulators do not currently meet the accuracy requirements for these cases.

We thus find that generating emulators for hybrid theories, such as HEFT, capable of modeling the real-space auto- and cross-spectra of galaxies and the matter field to $k\approx 0.6\,h\,\mathrm{Mpc}^{-1}$ with percent-level precision is easily achievable, even in models with 8 parameters.  The exact cost is driven by the range of the parameters that need to be sampled and the statistical precision required.  Even the most demanding cases, however, can be achieved with $\mathcal{O}(10^3)$ node-hours if cheap simulations are used.  If the same grid is run with higher resolution simulations (e.g.\ to measure additional statistics or for other science goals), the cost could rise as high as $10^{5-6}$ node-hours.

\section*{Acknowledgements}

The authors thank Nickolas Kokron for useful discussions regarding ZCV and for helpful comments on an earlier draft of the paper. AB and MW were supported by NASA ATP award 80NSSC24K0939.
This work was performed in part at Aspen Center for Physics, which is supported by National Science Foundation grant PHY-2210452.
This material is based upon work supported by the U.S. Department of Energy (DOE), Office of Science, Office of High-Energy Physics, under Contract No. DE–AC02–05CH11231, and by the National Energy Research Scientific Computing Center, a DOE Office of Science User Facility under the same contract.

\appendix
\section{Emulator construction}\label{app:emu_constr}

We use an emulator construction very similar to that used in ref.~\cite{DeRose_2023}. 
We use IR resummed $k$-expanded convolutional Lagrangian effective field theory predictions for $P_{\rm 1-loop}$ calculated using the RKECLEFT module of \texttt{velocileptors}. This choice is discussed further in Appendix \ref{app:ratio}. A third-order Savitsky-Golay filter of order 3 and window length 11 is applied to the ratio $P_{\rm model}/P_{\rm 1-loop}$ before emulation.

\begin{figure}
    \centering
    \includegraphics[width=\linewidth]{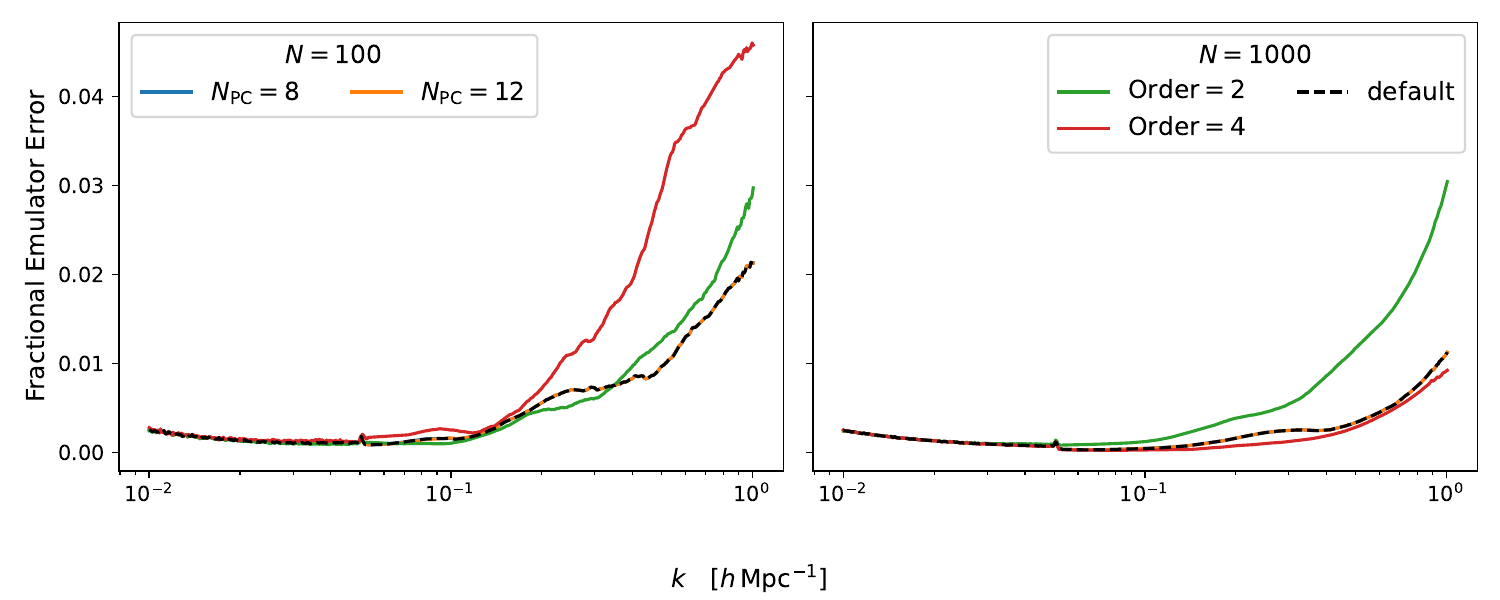}
    \caption{The $68$th percentile emulator errors for different choices of emulator hyperparameters. The left panel shows the error for an emulator trained with 100 training points, while the right panel shows the error for an emulator trained with 1000 points. The tier 2 parameter space is used for both. The curves in which just the number of principal components is varied lie under the black dashed curve.}
    \label{fig:hyperparam}
\end{figure}

As discussed in \ref{sec:Emulator}, the actual emulator uses a combination of polynomial chaos expansion (PCE) and principal component analysis (PCA). For PCE, we use order 3, and for PCA, we use 10 principal components. For further details of this implementation, we refer the reader to ref.~\cite{DeRose_2023}. Different choices for the PCE order and the number of principal components are compared in figure~\ref{fig:hyperparam}.  We see that the emulator error is highly stable under changes to the number of principal components. The PCE order has a more significant impact. Order 2 is a good choice for an emulator trained on 100 training points, where there is not enough data to properly constrain a large number of polynomial terms, but a poor choice for one trained on 1000 points where it is too `rigid'. Conversely, order 4 is a poor choice for an emulator trained on 100 points, but a good fit for one trained on 1000 points. Order 3 yields the most consistent results for the training dataset sizes we consider and is our default.

We have also investigated different choices for where to place the samples in our training set. Our fiducial choice is to use a Latin hypercube, but changing to Sobol or other low discrepancy sequences, as well as using different random seeds for these techniques, changes the accuracy of our emulators by significantly less than the variations observed when changing the number of points sampled.

\section{Basis spectra proxies}
\label{app:ratio}

\begin{figure}
    \centering
    \includegraphics[width=\linewidth]{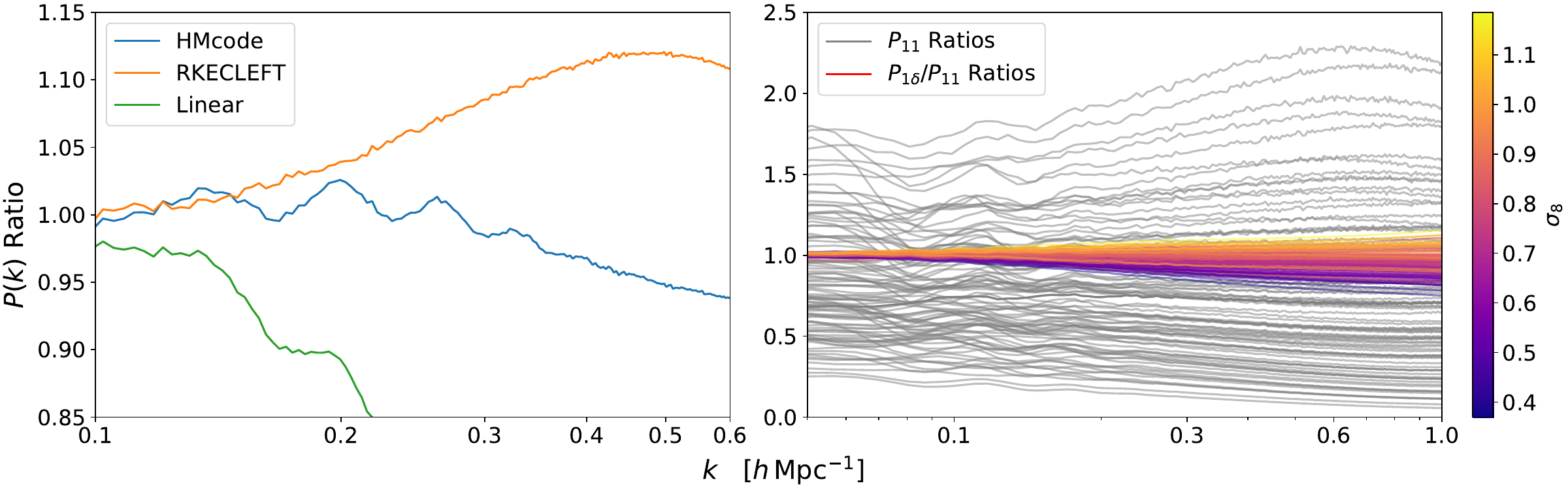}
    \caption{{\it Left}: Comparison of the dynamic ranges of the ratios of $P_{\rm N-body}$ as measured from the base Abacus simulations to the linear power spectrum, {\tt HMcode} predictions, and the 1-loop spectrum computed using RKECLEFT in \texttt{velocileptors}. The dynamic range of $P_{\rm lin}/P_{\rm N-body}$ is notably larger than the {\tt HMcode} and 1-loop ratios. BAO-like features are present in the {\tt HMcode} and linear power spectrum ratios, but not in the RKECLEFT 1-loop ratio.
    {\it Right}: Comparison of the ratios of matter power spectra (gray) and $P_{1\delta}/P_{11}$ for each Aemulus-$\nu$ simulation (colored). These ratios are taken with respect to measurements from box 91.
    All power spectra are computed or measured at $z=0$, which is the conservative choice for this inference.
    }
    \label{fig:RatioDynamicRange}
\end{figure}

It is common practice for emulators of the nonlinear matter power spectrum to emulate the nonlinear correction factor (NLC) $P_{\rm nonlinear}/P_{\rm lin}$ rather than $P_{\rm nonlinear}$ directly.  This is because an emulator performs better if the dynamical range of the quantity being emulated is reduced. However, ref.~\cite{EuclidEmulator_2021}, emulating $P_{\rm N-body}/P_{\rm lin}$, notes wiggles in emulator error on scales associated with BAO. We hypothesize that this is the result of dividing by $P_{\rm lin}$, rather than, e.g.\ $P_{\rm 1-loop}$, since linear theory does not properly account for the mode coupling induced by non-linear evolution that smooths the BAO oscillations. 

We expect that emulating ratios with a smaller dynamic range and that are relatively featureless yield smaller emulator error for a fixed training set size.  In figure~\ref{fig:RatioDynamicRange} (left), we compare the ratios $P_{\rm lin}/P_{\rm N-body}$, $P_{\rm HMcode}/P_{\rm N-body}$, and $P_{\rm 1-loop}/P_{\rm N-body}$, where $P_{\rm 1-loop}$ refers to RKECLEFT and $P_{N-{\rm body}}$ is the average of $P_{\rm mm}(k)$ for 10 of the Abacus base simulations with ZCV applied \cite{Maksimova_2021, Garrison_2021, Hadzhiyska_2021_CompaSO, Yuan_2021} and represents our best approximation to ``truth''. 

The dynamic range of $P_{\rm lin}/P_{\rm N-body}$ is significantly greater than that of $P_{\rm HMcode}/P_{\rm N-body}$. The dynamic range of $P_{\rm 1-loop}/P_{\rm N-body}$ is greater than that of $P_{\rm HMcode}/P_{\rm N-body}$ but less than $P_{\rm lin}/P_{\rm N-body}$.  We note the presence of wiggles in the HMcode ratio. These wiggles are also visible in the linear theory ratio, but are notably absent in the 1-loop ratio. This would imply that \texttt{velocileptors} is able to more accurately model the BAO wiggles and the broadening of the acoustic peaks resulting from nonlinear structure formation than {\tt HMcode}. As a result, though the 1-loop ratio dynamic range is larger than that of {\tt HMcode}, the featureless nature of $P_{\rm 1-loop}/P_{\rm N-body}$ suggests that one should emulate $P_{\rm N-body}/P_{\rm 1-loop}$ to optimize emulator performance.  For the HEFT spectra other than $P_{mm}$ we have no choice since there is no {\tt HMcode} version of those spectra.

These wiggles were indeed initially present in our emulator output when using {\tt HMcode} for $P_{\rm model}$ in Eq.~\ref{eqn:Gamma_defn}, which manifested as a series of wiggles around the BAO scale in our emulator error curves. Since we do not expect similar emulators trained on power spectra measured from $N$-body simulations to encounter this issue, as shown in the left panel of figure~\ref{fig:RatioDynamicRange}, we implement a smoothing spline over the data on these scales throughout this paper. 

The right panel of figure \ref{fig:RatioDynamicRange} shows that determining the emulator performance using $P_{11}$ provides an adequate proxy for the full HEFT modeling error. Physically this is because the dominant basis spectra, $P_{11}$, $P_{1\delta}$ and $P_{\delta \delta}$, scale with changes in cosmology quite similarly to $P_{\rm lin}$, times functions that are very smooth with cosmology. The right panel of figure \ref{fig:RatioDynamicRange} shows this, demonstrating that across our grid the spread in the ratios $P_{1\delta}/P_{11}$ vary significantly less than $P_{11}$.  While it is not shown explicitly in the figure, the same is true of $P_{\delta \delta}/P_{11}$.

The grey lines in \ref{fig:RatioDynamicRange} show the ratio of $P_{11}$ for each cosmology to that of $P_{11}$ in box 91 (whose cosmology lies near the middle of the Aemulus-$\nu$ parameter space).  The colored lines correspond to $P_{1\delta}/P_{11}$, also ``normalized'' by $P_{1\delta}/P_{11}$ in box 91. The colors of the different curves correspond to the values of $\sigma_8$ for each box. 
We see that the spread in the $P_{1\delta}/P_{11}$ ratios is significantly smaller than those in the $P_{11}$ ratios and in addition the amplitudes vary smoothly with cosmological parameters (in this case $\sigma_8$, which affects the degree of non-linearity and deviation from $P_{\rm lin}$ at high $k$). The small amplitude and smooth variation together suggest that emulating $P_{1\delta}$ (and $P_{\delta \delta}$) should not require additional training points if $P_{11}$ is emulated well.  The other basis spectra either behave similarly or are a much smaller contribution to the total power.

We also performed similar tests to check whether {\tt HMcode} provides an appropriate proxy for N-body simulation measurements. Again, the spread in $P_{\tt HMcode}$ with cosmology is significantly larger than the spread in $P_{{\rm Aem-}\nu}/P_{\tt HMcode}$. Given these tests, the addition of realistic noise to the HMcode spectra, and the smoothing spline over the BAO wiggles, we expect our HMcode-based emulators to provide an excellent proxy to N-body simulation measurements.

\section{ZCV for galaxy-matter cross-spectrum}

The previous, standard set of utilities, {\tt abacusutils}\footnote{Available at \href{https://github.com/abacusorg/abacusutils/tree/main}{https://github.com/abacusorg/abacusutils/tree/main}} included the ability to apply ZCV to the galaxy auto-power spectrum, but not the galaxy-matter cross-power spectrum. We have implemented ZCV for the cross-spectrum, and it is being included as part of the standard Abacus utilities. We note that ZCV has been used on galaxy cross-spectra previously \cite{Kokron_2022}, but has not been incorporated into the Abacus utilities until now.

\begin{figure}
    \centering
    \includegraphics[width=0.75\linewidth]{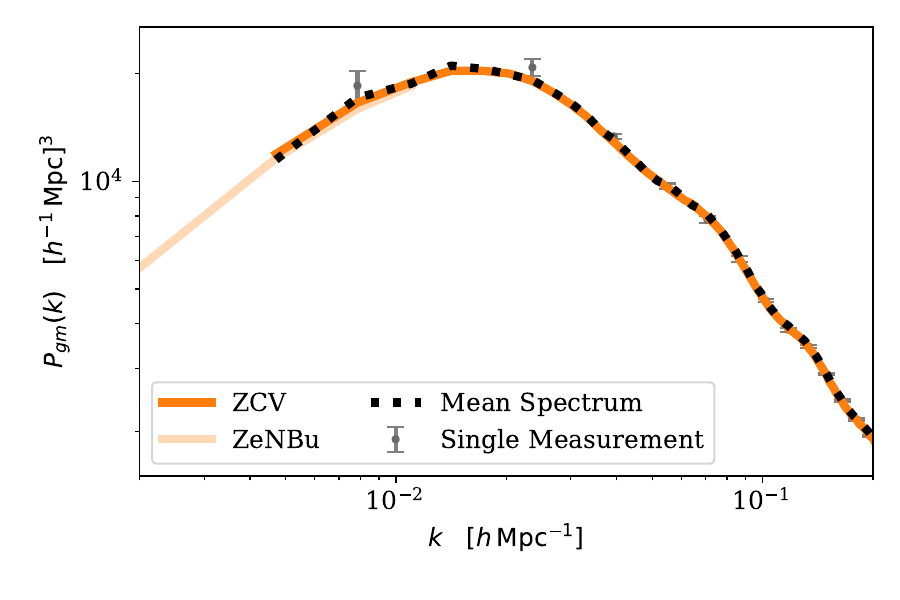}
    \caption{Measurements of the measured galaxy-matter cross-power spectrum with and without the use of Zel'dovich control variates. The points correspond to measurements from one base Abacus box, with error bars estimated from the ensemble of 13 base Abacus boxes. The dotted black line corresponds to the measured ensemble average of the same simulations. The light orange ZeNBu curve corresponds to the analytical Zel'dovich approximation predictions, and the opaque, dark orange curve corresponds to the galaxy cross-spectrum measured from a single base Abacus simulation with our ZCV code, which is being incorporated into the standard Abacus utilities.}
    \label{fig:ZCV_cross_check}
\end{figure}

The ZCV variance-reduced galaxy cross-spectrum is given by
\begin{equation}
    \hat{P}_{gm}(k) = \hat{P}_{gm}^{N}(k) - \beta^{\star}_{gm}(k) \left(\hat{P}_{gm}^{ZA}(k) - P_{gm}^{ZA}(k)\right)\,,
\end{equation}
where 
\begin{equation}
    \beta_{gm}^{\star}(k) = \frac{\mathrm{Cov}[P_{gm}^{N}, P_{gm}^{Z}]}{\mathrm{Var}[P_{gm}^{ZZ}]}\,,
\end{equation}
\begin{equation}
    {\rm Cov}[P_{gm}^{N}, P_{gm}^{Z}] \propto P_{gg}^{NZ}\,P_{mm}^{NZ} + P_{gm}^{NZ}\,P_{gm}^{ZN}\,,
\end{equation}
\begin{equation}
    {\rm Var}[P^{X}_{gm}] \propto P_{gg}^{X}P_{mm}^{X} + (P_{gm}^{X})^2,\quad X\in\{ZZ,NN\}\,,
\end{equation}
$\hat{P}_{gm}^{\rm ZA}(k)$ is the measured Zel'dovich spectrum, and $P_{gm}^{ZA}(k)$ can be computed analytically, or from multiple ZA measurements averaged together. All (co)variances are computed using the disconnected approximation. In the above expressions, 
\begin{equation}
    P_{gg}^{XY} = \sum_{\mathcal{O}_i,\mathcal{O}_j} b_{\mathcal{O}_i}b_{\mathcal{O}_j}\langle \mathcal{O}^X_i \mathcal{O}^Y_j \rangle\,, \qquad P_{gm}^{XY} = \sum_{\mathcal{O}_i} b_{\mathcal{O}}\langle \mathcal{O}^X_i \mathcal{O}^Y_{1_{cb}} \rangle\,, \quad \qquad P_{mm}^{XY} = \langle \mathcal{O}^X_{1_{cb}} \mathcal{O}^Y_{1_{cb}}\rangle\,,
\end{equation}
and $\mathcal{O}_i \in \{\mathbf{1}_{cb},\delta,:\!\!\delta^2\!\!:,:\!\!s^2\!\!:,\nabla^2\delta\}\,$.

\bibliographystyle{JHEP}
\bibliography{JHEP}
\end{document}